\documentclass[a4paper,11pt]{article}

\pdfoutput 1

\usepackage{jheppub}

\usepackage[utf8]{inputenc}
\usepackage[T1]{fontenc}
\usepackage{lmodern}
\usepackage{textcomp}
\usepackage{microtype}

\usepackage{xspace}
\newcommand{\software}[2][]{\texttt{#2}\xspace#1}

\usepackage{units}
\usepackage{cases}

\allowdisplaybreaks[1]

\DeclareMathOperator{\SU}{SU}
\DeclareMathOperator{\U}{U}
\DeclareMathOperator{\SO}{SO}
\DeclareMathOperator{\tr}{tr}
\DeclareMathOperator{\BR}{BR}

\newcommand{\abs}[1]{\left\lvert#1\right\rvert}
\newcommand{\order}[1]{\mathcal O\left(#1\right)}
\newcommand{\at}[2]{\left.#1\right\rvert_{#2}}
\newcommand{\like}[2]{L\!\left(#1\middle\vert#2\right)}
\newcommand{\sig}[2]{Z\!\left(#1\middle\vert#2\right)}
\newcommand{\Mass}{\mathcal M}
\newcommand{\Lag}{\mathcal L}

\newcommand{\field}[1]{\ensuremath{{#1}}\xspace}

\newcommand{\gaugeR}{\field{u_3^c}}
\newcommand{\GaugeR}{\field{U^c}}
\newcommand{\gaugeL}{\field{q_3}}
\newcommand{\GaugeL}{\field{U}}
\newcommand{\gauge}{c}
\renewcommand{\Gauge}{\widehat c}

\newcommand{\massR}{\field{t^{\prime c}}}
\newcommand{\MassR}{\field{T^{\prime c}}}
\newcommand{\massL}{\field{t^\prime}}
\newcommand{\MassL}{\field{T^\prime}}

\newcommand{\breakR}{\field{t^c}}
\newcommand{\BreakR}{\field{T^c}}
\newcommand{\breakL}{\field{t}}
\newcommand{\BreakL}{\field{T}}

\newcommand{\breakQuad}{a}
\newcommand{\breakCube}{b}

\usepackage[font=small]{caption}
\usepackage[font=footnotesize]{subcaption}
\usepackage{booktabs}
\usepackage{multirow}

\graphicspath{{figs/}}

\makeatletter\g@addto@macro\bfseries{\boldmath}\makeatother

\newcommand*{\doi}[1]{\href{http://dx.doi.org/#1}{doi: #1}}
\newcommand{\No}[1]{\textnumero~#1}

\usepackage{etoolbox}
\DeclareRobustCommand{\mathlist}[1]{%
\def\nextitem{\gdef\nextitem{,\:}}%
\renewcommand*{\do}[1]{ \nextitem{##1}}%
\docsvlist{#1}%
}
\newcommand{\pair}[2]{\ensuremath{\left(\mathlist{#1,#2}\right)}}
\newcommand{\listunit}[2][]{\unit[\mathlist{#1}]{#2}}

\newcommand{\graphic}[2][1]{\includegraphics[width=#1\textwidth]{#2}}

\usepackage{array}
\newcommand{\undefcolumntype}[1]{\expandafter\let\csname NC@find@#1\endcsname\relax}
\newcommand{\forcenewcolumntype}[1]{\undefcolumntype{#1}\newcolumntype{#1}}
\AtBeginDocument{\forcenewcolumntype{|}{@{\hskip\tabcolsep}}}

\title{Testing Naturalness}

\author[a]{Chuan-Ren Chen,}
\emailAdd{crchen@ntnu.edu.tw}
\affiliation[a]{Department of Physics, National Taiwan Normal University, Taipei 116, Taiwan}

\author[b,c]{Jan Hajer,}
\emailAdd{jan@hajer.com}
\affiliation[b]{Institute for Advanced Studies, The Hong Kong University of  Science and Technology, Clear Water Bay, Kowloon, Hong Kong S.A.R, China}

\author[c]{Tao Liu,}
\emailAdd{taoliu@ust.hk}
\affiliation[c]{Department of Physics, The Hong Kong University of  Science and Technology, Clear Water Bay, Kowloon, Hong Kong S.A.R, China}

\author[d,e]{Ian Low}
\emailAdd{ilow@northwestern.edu}
\affiliation[d]{Department of Physics and Astronomy, Northwestern University, Evanston, IL 60208, USA}
\affiliation[e]{High Energy Physics Division, Argonne National Laboratory, Argonne, IL 60439, USA}

\author[f]{and Hao Zhang}
\emailAdd{zhanghao@ihep.ac.cn}
\affiliation[f]{Institute of High Energy Physics, Chinese Academy of Sciences, Beijing 100049, China}

\abstract{%
Solutions to the electroweak hierarchy problem typically introduce a new symmetry to stabilize the quadratic ultraviolet sensitivity in the self-energy of the Higgs boson.
The new symmetry is either broken softly or collectively, as for example in supersymmetric and  little Higgs theories.
At low energies such theories  contain naturalness partners of the Standard Model fields which are responsible for canceling the quadratic divergence in the squared Higgs mass.
Post the discovery of any partner-like particles, we propose to test  the aforementioned cancellation  by measuring  relevant Higgs couplings.
Using the fermionic top partners in little Higgs theories as an illustration, we  construct a simplified model for naturalness and initiate a study on testing naturalness.
After electroweak symmetry breaking, naturalness in the top sector requires  $\breakQuad_\BreakL = - \lambda_\breakL^2$ at leading order, where $\lambda_\breakL$ and $\breakQuad_\BreakL$ are the Higgs couplings to a pair of top quarks and top partners, respectively.
Using a multivariate method of Boosted Decision Tree to tag boosted particles in the Standard Model, we show that, with a luminosity of $\unit[30]{ab^{-1}}$ at a \unit[100]{TeV} $pp$-collider, naturalness could be tested with a precision of \unit[10]{\%} for a top partner mass up to 2.5 TeV.
}

\begin{document}

\maketitle

\section{Introduction}

The naturalness problem~\cite{'tHooft:1979bh}, motivated by the smallness of the Standard Model~(SM) particle masses in comparison to a UV cutoff, has been the driving force behind theoretical developments in particle physics for several decades.
Especially after the discovery of the 125 GeV Higgs boson at the Large Hadron Collider (LHC) in 2012~\cite{Aad:2012tfa, Chatrchyan:2012ufa}, there is a renewed urgency to understand why its mass is so much smaller than the next known scale in particle physics, the Planck scale, given the quadratic ultraviolet sensitivity in the self-energy of the Higgs.
In the past decades, many theories of naturalness have been proposed.
Typically a new symmetry which protects the squared mass of the SM-like Higgs boson from quadratic UV-sensitivity is introduced.
Two eminent examples are supersymmetry~\cite{Haber:1984rc} and spontaneously broken global symmetry~\cite{Kaplan:1983fs, Kaplan:1983sm}, which require the existence of naturalness partners coupling to the Higgs boson.
The symmetry-protected relation among the Higgs couplings with the SM particles and their partners, called the naturalness sum rule, ensures cancellation of the quadratic divergence in the squared Higgs mass.
Confirming or refuting theories of naturalness is one of the most important goals for the current and future high energy collider experiments.

Among the naturalness partners, the partner of the SM top quark typically plays the most prominent role.
Currently, a tremendous amount of effort at the LHC has been going into searches for top partners~\cite{1434358, Aaboud:2016lwz, ATLAS:2016sno, CMS:2016ete}, whose discovery potential at a next-generation hadron collider has also been studied~\cite{Golling:2016gvc}.
However, the discovery of a partner-like particle is only the first step in  confirming theories of naturalness.
In order to ensure that the discovered partner-like particle is not a naturalness impostor, but indeed the one that yields the cancellation of the quadratic UV-sensitivity in the Higgs self-energy, we need to further measure the naturalness sum rule.
In this article, we propose to probe the naturalness sum rule at hadron colliders, as a crucial step to test theories of naturalness.
For convenience, we rephrase the naturalness sum rule $\at{\Delta m_H^2}{\text{SM}} + \at{\Delta m_H^2}{\text{NP}} = 0$, which relates the different contributions to the quadratically divergent quantum corrections to the squared Higgs mass $\Delta m_H^2$, into a naturalness parameter
\begin{align}
    \mu
 &= - \frac{\at{\Delta m_H^2}{\text{NP}}}{\at{\Delta m_H^2}{\text{SM}}}
\ ,
 &  \at{\mu}{\text{naturalness}}
  = 1
\ . \label{naturalness parameter}
\end{align}

Note that in the SM the top quarks contribute the strongest to the UV-sensitivity of the Higgs self-energy.
As a first illustration for testing naturalness, we will analyze fermionic top partners which carry QCD colors, as predicted by little Higgs models~\cite{ArkaniHamed:2001nc, ArkaniHamed:2002qx, ArkaniHamed:2002qy}.%
\footnote{%
The case of scalar top partners as predicted by e.g. supersymmetry will be pursued in a future work.
}
Similar colored top partners exist in the so-called composite Higgs models as well~\cite{Contino:2003ve, Agashe:2004rs}.
We define a simplified model consisting of the SM top fields and a pair of vector-like fermionic top-partners, which are either singlets or doublets under $\SU(2)_w$.
Then we show that the naturalness sum rule leads to an especially simple form after electroweak symmetry breaking, at leading order in the Higgs vacuum expectation value (VEV)
\begin{equation}
    \breakQuad_\BreakL
  = -\abs{\lambda_\breakL}^2
  + \order{\frac{v^2}{m_\BreakL^2}}
\ , \label{naturalness}
\end{equation}
here $m_\BreakL$ is the top partner mass and the relation involves at leading order only the couplings of the Higgs to a pair of top quarks $\lambda_\breakL$ and a pair of top-partners $\breakQuad_\BreakL$.
This suggests that, in order to test the naturalness sum rule up to $\order{{v^2}/{m_\BreakL^2}}$, only two measurements are necessary, one for each of the couplings $\lambda_\breakL$ and $\breakQuad_\BreakL$.
We then  proceed to study how this relation can be measured to quantify the degree of cancellation in the contribution of the top sector to the quadratic divergence of the Higgs self energy in a future \unit[100]{TeV} hadron collider~\cite{FCC, CEPC-SPPCStudyGroup:2015csa}.

In Section~\ref{sec:simplified model} we introduce a simplified model for naturalness, by extending the SM top sector with a pair of vector-like top partners, and derive relation from the requirement of naturalness.
Based on this discussion we design a detector study dedicated to test naturalness of fermionic top partners at a \unit[100]{TeV} collider in Section~\ref{sec:analysis}.
Finally we conclude in Section~\ref{sec:conclusion}.

\section{A Simplified Model for Naturalness}\label{sec:simplified model}

In the SM, the UV-sensitivity of the Higgs self-energy originates from three major sources which, in order of decreasing magnitude, are the coupling to top quarks, electroweak gauge bosons and the Higgs boson itself.
In a typical natural theory, the contributions of top quarks, EW gauge bosons and the Higgs boson are expected to be canceled by their respective partner particles.
In this work we consider a top sector with fermionic partner particles, which carry QCD colors and could be produced copiously in a hadron collider, if their masses are not too heavy.
These top partners are vector-like and massive, thus allowing a quartic coupling with $H^\dagger H$ which enables the cancellation of the top quadratic divergence in the Higgs self-energy.
We focus on the cases where the vector-like top partners carry the same EW charges as either the right-chiral top quark~\gaugeR~\cite{ArkaniHamed:2002qy, Cheng:2005as} or the left-chiral top quark $\gaugeL = \pair{d_3}{u_3}$~\cite{Chang:2003zn}.
In the case where the vector-like pair \pair{\GaugeR}{\GaugeL} is a singlet under $\SU(2)_w$ the effective Lagrangian for the non-linear coupling of the top sector to the Higgs scalar~$H$ can be expanded to
\begin{multline}
    \Lag_\GaugeL
= \gaugeR
    \left(
        \gauge_0 f \GaugeL
      + \gauge_1 H^\dagger \gaugeL
      + \frac{\gauge_2}{2f} \abs{H}^2 \GaugeL
      + \frac{\gauge_3}{6f^2} \abs{H}^2 H^\dagger \gaugeL
      + \dots
    \right) \\
  + \GaugeR
    \left(
        \Gauge_0 f \GaugeL
      + \Gauge_1 H^\dagger \gaugeL
      + \frac{\Gauge_2}{2f} \abs{H}^2 \GaugeL
      + \frac{\Gauge_3}{6f^2} \abs{H}^2 H^\dagger \gaugeL
      + \dots
    \right)
  + \text{h.c.}
\ . \label{eq:gauge Lagrangian}
\end{multline}
The mass dimension in these couplings is given by the symmetry breaking scale~$f$ and we imply $\abs{H}^2 = H^\dagger H$.
The parameter $\gauge_i$ and $\Gauge_i$ are free and for odd-dimensional couplings they can be complex.
A linear combination of \gaugeR and \GaugeR together with \GaugeL becomes massive with a top partner mass of
\begin{align}
    m_\MassL
 &= f \gauge
\ ,
  & \gauge
 &= \sqrt{\gauge_0^2 + \Gauge_0^2}
\ . \label{eq:partner mass}
\end{align}
Whereas, the orthogonal combination remains massless and acquires a Yukawa coupling to the Higgs doublet.
In terms of these mass eigenstates
\begin{subequations}
\label{eq:mass rotation}
\begin{align}
    \massR
 &= \frac{\Gauge_0 \gaugeR - \gauge_0 \GaugeR}{\gauge}
\ ,
 & \massL
 &= \gaugeL
\ ,
 \\ \MassR
 &= \frac{\Gauge_0 \GaugeR + \gauge_0 \gaugeR}{\gauge}
\ ,
 & \MassL
 &= \GaugeL
\ ,
\end{align}
\end{subequations}
the Lagrangian~\eqref{eq:gauge Lagrangian} becomes
\begin{multline}
    \Lag_\MassL
  = m_\MassL \MassR \MassL
  + \lambda_\massL H^\dagger \massR \massL
  + \lambda_\MassL H^\dagger \MassR \massL
  + \frac{\alpha_\massL}{2 m_\MassL} \abs{H}^2 \massR \MassL
  + \frac{\alpha_\MassL}{2 m_\MassL} \abs{H}^2 \MassR \MassL \\
  + \frac{\beta_\massL}{6 m_\MassL^2} \abs{H}^2 H^\dagger \massR \massL
  + \frac{\beta_\MassL}{6 m_\MassL^2} \abs{H}^2 H^\dagger \MassR \massL
  + \order{H^4}
  + \text{h.c.}
\ , \label{eq:mass Lagrangian}
\end{multline}
where the couplings are given by
\begin{subequations}
\label{eq:parameter}
\begin{align}
   \lambda_\massL
 &= \frac{\Gauge_0 \gauge_1 - \gauge_0 \Gauge_1}{\gauge}
\ ,
  & \lambda_\MassL
 &= \frac{\gauge_0 \gauge_1 + \Gauge_0 \Gauge_1}{\gauge}
\ , \label{eq:Yukawa couplings}
 \\ \alpha_\massL
 &= \Gauge_0 \gauge_2 - \gauge_0 \Gauge_2
\ ,
  & \alpha_\MassL
 &= \gauge_0 \gauge_2 + \Gauge_0 \Gauge_2
\ , \label{eq:quadratic couplings}
 \\
    \beta_\massL
 &= \left(\Gauge_0 \gauge_3 - \gauge_0 \Gauge_3 \right) \gauge
\ ,
  & \beta_\MassL
 &= \left(\gauge_0 \gauge_3 + \Gauge_0 \Gauge_3 \right) \gauge
\ . \label{eq:cubic couplings}
\end{align}
\end{subequations}

The case in which the vector-like pair \pair{Q^c}{Q} consists of $\SU(2)_w$ doublets can be written as
\begin{multline}
    \Lag_Q
= \left(
        Q^c \gauge_0 f
      + \gaugeR \gauge_1 H^\dagger
      + Q^c \frac{\gauge_2}{2f} \abs{H}^2
      + \gaugeR \frac{\gauge_3}{6f^2} \abs{H}^2 H^\dagger
      + \dots
    \right) \gaugeL \\
 + \left(
        Q^c \Gauge_0 f
      + \gaugeR \Gauge_1 H^\dagger
      + Q^c \frac{\Gauge_2}{2f} \abs{H}^2
      + \gaugeR \frac{\Gauge_3}{6f^2} \abs{H}^2 H^\dagger
      + \dots
    \right) Q
  + \text{h.c.}
\ . \label{eq:doublet Lagrangian}
\end{multline}
The rotation into mass eigenstates leads to a Lagrangian identical to~\eqref{eq:mass Lagrangian} with the replacement of
\begin{align}
    \MassR
  &\leftrightarrow
    \MassL
\ ,
  & \massR
  &\leftrightarrow
    \massL
\ .
\end{align}
Although the following discussion is based on Lagrangian~\eqref{eq:gauge Lagrangian} it can also be applied to Lagrangian~\eqref{eq:doublet Lagrangian} and our results can easily be adjusted to correspond to that case.

The contribution of Lagrangian~\eqref{eq:mass Lagrangian} to the Higgs potential can be computed using the Coleman-Weinberg potential~\cite{Coleman:1973jx} and the quadratically divergent part is
\begin{align}
    V^T_\text{quad}
 &= \frac{\Lambda^2}{16\pi^2} \tr \Mass^2
\ ,
  & \Mass^2
 &= \Mass(H)^\dagger \Mass(H)
\ ,
\end{align}
where $\Mass(H)$ is the fermion mass matrix with the Higgs field $H$ as a background field, and $\Lambda$ is the cutoff scale of the theory.
In the basis \pair{\massR}{\MassR} and \pair{\massL}{\MassL} the mass matrix reads up to quadratic order in the Higgs field
\begin{equation}
    \Mass(H)
  = \begin{pmatrix}
      0
    & 0
   \\ 0
    & m_\MassL
   \end{pmatrix}
 + \begin{pmatrix}
      \lambda_\massL
    & 0
   \\ \lambda_\MassL
    & 0
   \end{pmatrix}
   H
 + \begin{pmatrix}
      0
    & \alpha_\massL
   \\ 0
    & \alpha_\MassL
   \end{pmatrix}
   \frac{\abs{H}^2}{2m_\MassL}
 + \order{H^3}
\ . \label{eq:mass matrix}
\end{equation}
The absence of quadratically divergent contributions in the Higgs mass
\begin{equation}
    \Delta m^2_\text{quad}
  = \frac{1}{2} \at{\frac{\partial^2 V^T_\text{quad}}{\partial \abs{H}^2}}{H=0}
  \equiv
    0
\ , \label{eq:quadratic divergent}
\end{equation}
is then equivalent to the requirement that the coefficient of the $\abs{H}^2$ term in $\tr \Mass^2$ vanishes
\begin{equation}
    \tr \Mass^2
  = \text{constant}
  + \order{H^3}
\ .
\label{eq:cancelation}
\end{equation}
Several comments follow from this equation.
First, due to the structure of the matrix~\eqref{eq:mass matrix}, no linear term and hence no tadpole contribution arises.
Second, $\alpha_\massL$ only contributes to $\abs{H}^4$ terms and does not enter into the cancellation of quadratic divergences in the Higgs mass.
Finally, requiring the coefficients of the $\abs{H}^2$ term to vanish leads to the naturalness sum rule
\begin{equation}
    \alpha_\MassL
  = - \abs{\lambda_\MassL}^2
  - \abs{\lambda_\massL}^2
\ . \label{eq:naturalness sum rule}
\end{equation}
It is worthwhile pointing out that the $\lambda_\MassL$ term contributes to the quadratic divergences with the same sign as the SM top quark, and the cancellation hinges entirely on the existence of the four-point coupling $\alpha_\MassL$ which must have a negative sign. In terms of the parameter in the Lagrangians of the gauge eigenstates~\eqref{eq:gauge Lagrangian} and~\eqref{eq:doublet Lagrangian} this naturalness sum rule reads
\begin{equation}
  \abs{\gauge_1}^2
  + \gauge_0 \gauge_2
  =- \abs{\Gauge_1}^2
  - \Gauge_0 \Gauge_2
\ . \label{naturalness restriction 2}
\end{equation}

\begin{table}
\centering
\begin{tabular}{crccccccccc}
    \toprule
    Model \hfill Coset
  &
  & $\SU(2)$
  & $\gauge_0$
  & $\gauge_1$
  & $\gauge_2$
  & $\Gauge_0$
  & $\Gauge_1$
  & $\Gauge_2$
 \\ \midrule
    Toy model \hfill $\frac{\SU(3)}{\SU(2)}$
  &\cite{Perelstein:2003wd}
  & \textbf 1
  & $\lambda_1$
  & $-\lambda_1$
  & $-\lambda_1$
  & $\lambda_2$
  & 0
  & 0
 \\ Simplest \hfill $\left(\frac{\SU(3)}{\SU(2)}\right)^2$
  &\cite{Kaplan:2003uc}
  & \textbf 1
  & $\lambda$
  & $-\lambda$
  & $-\lambda$
  & $\lambda$
  & $\lambda$
  & $-\lambda$
 \\ Littlest Higgs \hfill $\frac{\SU(5)}{\SO(5)}$
  &\cite{ArkaniHamed:2002qy}
  & \textbf 1
  & $\lambda_1$
  & $-\sqrt2i\lambda_1$
  & $-2\lambda_1$
  & $\lambda_2$
  & 0
  & 0
 \\ Custodial \hfill $\frac{\SO(9)}{\SO(5)\SO(4)}$
  &\cite{Chang:2003zn}
  & \textbf 2
  & $y_1$
  & $\frac{i}{\sqrt 2} y_1$
  & $-\frac{1}{2} y_1$
  & $y_2$
  & 0
  & 0
 \\ \midrule
    $T$-parity invariant \hfill $\frac{\SU(3)}{\SU(2)}$
  &\cite{Cheng:2005as}
  & \textbf 1
  & $\lambda$
  & $-\lambda$
  & $-\lambda$
  & $-\lambda$
  & $-\lambda$
  & $\lambda$
 \\ $T$-parity invariant \hfill $\frac{\SU(5)}{\SO(5)}$
  &\cite{Cheng:2005as}
  & \textbf 1
  & $\lambda$
  & $-\sqrt2i\lambda$
  & $-2\lambda$
  & $-\lambda$
  & $-\sqrt2i\lambda$
  & $2\lambda$
 \\ \midrule
  Mirror twin Higgs \hfill $\frac{\SU(4)\U(1)}{\SU(3)\U(1)}$
  &\cite{Burdman:2014zta}
  & \textbf 1
  & 0
  & $i \lambda_\breakL$
  & 0
  & $\lambda_\breakL$
  & 0
  & $-\lambda_\breakL$
 \\ \bottomrule
\end{tabular}
\caption{%
Parameter of the couplings between top quarks, top partners and the Higgs scalar in the gauge eigenstate basis as defined in Lagrangian~\eqref{eq:gauge Lagrangian} for some representative models.
In the column labeled $\SU(2)$ we indicate if the top partners are singlets or embedded into doublets under $\SU(2)_w$ and hence if this model corresponds to the Lagrangian~\eqref{eq:gauge Lagrangian} or~\eqref{eq:doublet Lagrangian}.
}
\label{tab:models}
\end{table}

The best known examples of models described by Lagrangian~\eqref{eq:mass Lagrangian} and subject to the naturalness sum rule~\eqref{eq:naturalness sum rule} are little Higgs models.
Some typical examples are presented in Table~\ref{tab:models}.%
\footnote{%
In Table~\ref{tab:models} we have also considered mirror twin Higgs models~\cite{Burdman:2014zta}, although their phenomenology differs significantly from the one in little Higgs theories under consideration.
See also the discussion at the end of this section.
}
The collective symmetry breaking in the Yukawa sector of little Higgs models yields the relations
\begin{align}
    \abs{\gauge_1}^2
 &= - \gauge_0 \gauge_2
\ ,
  & \abs{\Gauge_1}^2
 &= - \Gauge_0 \Gauge_2
\ ,
\end{align}
as can easily be checked in Table~\ref{tab:models}.
This ensures at one-loop level the cancellation of the quadratic divergences from the top sector according to the condition \eqref{naturalness restriction 2}.
A common but by no means necessary additional restriction is to couple only one of the top partners to the Higgs field by setting $\Gauge_1 = \Gauge_2 = \dots = 0$, which leads to
\begin{equation}
    \frac{\beta_\massL}{\beta_\MassL}
  = \frac{\alpha_\massL}{\alpha_\MassL}
  = \frac{\lambda_\massL}{\lambda_\MassL}
  = \frac{\Gauge_0}{\gauge_0}
\ . \label{eq:additional restriction}
\end{equation}
For phenomenological reasons it can be desirable to introduce $T$-parity~\cite{Cheng:2003ju,Cheng:2004yc}.
When $T$-parity is implemented in the top sector~\cite{Cheng:2005as} the top partner particles are $T$-odd, while SM particles are $T$-even.
Therefore the mixing terms have to vanish $\beta_\MassL = \alpha_\massL  = \lambda_\MassL = 0$.
Before rotation into the mass eigenstates this corresponds to the condition $\Gauge_i = \mp \gauge_i$ for even and odd $i$, respectively.

Finally, expanding the Higgs field around its VEV%
\footnote{%
In theories where the Higgs arises as a pseudo-Nambu-Goldstone boson from a spontaneously broken symmetry, the VEV of the Higgs differs slightly from the SM VEV of \unit[174]{GeV}.
The difference is formally of order of $\nicefrac{v^2}{m_\BreakL^2}$, similar to terms that have been neglected during the derivation of the naturalness sum rule.
}
\begin{align}
    H
 &= \pair{0}{v}^T
  + \frac1{\sqrt{2}} \pair{h^+}{h}^T
\ ,
  & v
 & \approx \unit[174]{GeV}
\ ,
\end{align}
and rotating the fermion fields to linear order in $v$ into mass eigenstates
\begin{subequations}
\label{eq:electroweak rotation}
\begin{align}
    \breakR
 &= \massR
  + \order{\frac{v^2}{m_\MassL^2}}
\ ,
 & \breakL
 &= \massL
  - \lambda_\MassL^* \frac{v}{m_\MassL} \MassL
  + \order{\frac{v^2}{m_\MassL^2}}
\ ,
 \\ \BreakR
 &= \MassR
  + \order{\frac{v^2}{m_\MassL^2}}
\ ,
 & \BreakL
 &= \MassL
  + \lambda_\MassL \frac{v}{m_\MassL} \massL
  + \order{\frac{v^2}{m_\MassL^2}}
\ ,
\end{align}
\end{subequations}
leads to the Lagrangian
\begin{multline}
    \Lag_\BreakL
  = m_\BreakL \BreakR\BreakL
  + \lambda_\breakL v \breakR \breakL
  + \frac{\lambda_\breakL}{\sqrt 2} h \breakR \breakL
  + \frac{\lambda_\BreakL}{\sqrt 2} h \BreakR \breakL
  + \frac{\breakQuad_\breakL v}{\sqrt 2 m_\BreakL} h \breakR \BreakL
  + \frac{\breakQuad_\BreakL v}{\sqrt 2 m_\BreakL} h \BreakR\BreakL
\\+ \frac{\alpha_\breakL}{4 m_\BreakL} h^2 \breakR \BreakL
  + \frac{\alpha_\BreakL}{4 m_\BreakL} h^2 \BreakR\BreakL
  + \frac{\breakCube_\breakL v}{4 m_\BreakL^2} h^2 \breakR \breakL
  + \frac{\breakCube_\BreakL v}{4 m_\BreakL^2} h^2 \BreakR \breakL
  + \order{h^3, \frac{v^2}{m_\BreakL^2}}
  + \text{h.c.}
\ . \label{eq:electroweak Lagrangian}
\end{multline}
The couplings $m_\BreakL$, $\lambda_{\breakL,\BreakL}$ and $\alpha_{\breakL,\BreakL}$ are up to linear order in $\nicefrac{v}{m_\BreakL}$ identical to their primed counterparts, and we have collected the higher order correction in Appendix~\ref{sec:corrections}.
The new VEV induced couplings are
\begin{subequations}
\label{eq:electroweak parameter}
\begin{align}
    \breakQuad_\breakL
 &= \alpha_\massL
  + \lambda_\MassL^* \lambda_\massL
\ ,
  & \breakQuad_\BreakL
 &= \alpha_\MassL
  + \abs{\lambda_\MassL}^2
\ , \label{eq:electroweak Yukawa}
 \\  \breakCube_\breakL
 &= \beta_\massL
  - \alpha_\massL \lambda_\MassL
\ ,
  & \breakCube_\BreakL
 &= \beta_\MassL
  - \alpha_\MassL \lambda_\MassL
\ . \label{eq:electroweak quadratic}
\end{align}
\end{subequations}
The naturalness sum rule~\eqref{eq:naturalness sum rule} leads at leading order in $v$ to the remarkable relation
\begin{equation}
\boxed{
    \breakQuad_\BreakL
  = -\abs{\lambda_\breakL}^2
  + \order{\frac{v^2}{m_\BreakL^2}}
}
\ , \label{eq:electroweak naturalness sum rule}
\end{equation}
which has significant implications on tests of the naturalness sum rule at colliders.

Naively, the relation~\eqref{eq:naturalness sum rule} suggests that three measurements, one for each of the couplings $\lambda_\massL$, $\lambda_\MassL$ and $\alpha_\MassL$, are necessary in order to test naturalness.
However, as $\lambda_\massL$, $\lambda_\MassL$ and $\alpha_\MassL$ are not defined in the basis of mass eigenstates after electroweak symmetry breaking, the separate reconstructing of their values becomes rather challenging.
Previously it was suggested~\cite{Perelstein:2003wd} to test the naturalness relation~\eqref{eq:naturalness sum rule} by measuring the $\lambda_\MassL$ coupling via the top partner decay $\BreakL\to \breakL h$, and extracting the information of $\alpha_\MassL$ with the relation
\begin{equation}
    \alpha_\MassL
  = \lambda_\MassL \frac{m_\MassL}{f}
\ . \label{assumption1}
\end{equation}
However, the proposed strategy is limited by two observations.
First this relation is not generic for all little Higgs models.
The $\lambda_\MassL$ term for example can be eliminated by imposing $T$-parity in the top sector~\cite{Cheng:2005as}.
Second, it implies that in order to extracting the value of the coupling $\alpha_\MassL$ it is necessary to measure the decay constant~$f$.
In contrast, the naturalness sum rule~\eqref{eq:electroweak naturalness sum rule} has the virtue of being generic for all models which can be described by the simplified model under consideration.
Especially it is independent of the assumption on $T$-parity.
It is exact up to order $\nicefrac{v}{m_\BreakL}$, and involves only two free parameters, the coupling of the  Higgs to two top quarks and to two top partners.
Notably, this does not involve the measurement of the decay constant~$f$.
In conclusion the relation~\eqref{eq:electroweak naturalness sum rule} greatly simplifies testing the naturalness sum rule.

Although, the naturalness sum rules \eqref{eq:naturalness sum rule}, \eqref{naturalness restriction 2} and \eqref{eq:electroweak naturalness sum rule} are derived for vector-like top partners in little Higgs models, they can also be applied to mirror twin Higgs models~\cite{Burdman:2014zta}.
In this case, top quarks carry QCD charge while top partners carry mirror QCD charge, which forbids mixing between them.
This is ensured by the condition that $\gauge_0 = \Gauge_1 = \gauge_2 \equiv 0$, which yields the relations $\beta_\MassL = \alpha_\massL  = \lambda_\MassL = 0$.
Note the similarity to the case of little Higgs models with $T$-parity.
However, in this case the naturalness sum rule \eqref{naturalness restriction 2} is reduced to
\begin{equation}
    \abs{\gauge_1}^2
  = - \Gauge_0 \Gauge_2
\ .
\end{equation}
Although, the collider kinematics of colorless top partners differs significantly from that of colored top partners, we list twin Higgs models as one possibility in Table~\ref{tab:models}.
However, the collider analysis in the following section applies only to models with colored top partners such as little Higgs models.

\section{Collider Analysis}\label{sec:analysis}

Based on the discussions above, the measurements of the couplings of a Higgs with a pair of top quarks and a pair of top partners constitute the two cornerstones for testing the naturalness sum rule.
The former measurement has been extensively discussed in the literature, for searches at the LHC~\cite{Drollinger:2001ym, Plehn:2009rk, Artoisenet:2013vfa, Khachatryan:2014qaa, Aad:2014lma, Khachatryan:2015ila}, at a future $e^+e^-$ collider~\cite{Weiglein:2004hn, Klute:2013cx, Moortgat-Picka:2015yla}, and at a \unit[100]{TeV} hadron collider~\cite{Plehn:2015cta}.
We will focus on the measurement of the Higgs coupling to two top partners~$\breakQuad_\BreakL$ via $\BreakR\BreakL h$ production at a \unit[100]{TeV} hadron collider.
The naturalness parameter~\eqref{naturalness parameter} applied to this analysis reads
\begin{equation}
    \mu_\breakL
  = - \frac{\breakQuad_\BreakL}{\lambda_\breakL^2}
  + \order{\frac{v^2}{m_\BreakL^2}}
\ .
\end{equation}
If the naturalness sum rule~\eqref{eq:naturalness sum rule} is satisfied $\mu_\breakL$ is up to an order of $\nicefrac{v^2}{m_\BreakL^2}$ equal to one.
The analysis proposed here is blind to the sign of~$\breakQuad_\BreakL$.
We leave the removal of this degeneracy for a future publication.

\subsection{Simulation of Signal and Background}\label{sec:simulation}

In little Higgs models, fermionic top partners decay mainly into heavy quarks and bosons $B = \mathlist{W^\pm,Z,h}$.
The equivalence theorem ensures that in the limit $f \gg v$ the branching ratios~(BR) relate according to~\cite{delAguila:1989rq}
\begin{equation}
    \BR(T\to th)
  \simeq
     \BR(T\to tZ)
  \simeq
    \frac{1}{2} \BR(T\to Wb)
  \simeq
    \unit[25]{\%}
\ . \label{eq:branching ratios}
\end{equation}
This relation receives a correction of order $\nicefrac{v}{m_\BreakL}$ from the $\breakQuad_\breakL$ term in
\begin{align}
    \Lag_\BreakL
 &\supset
    \frac{1}{\sqrt 2}
    \left(
        \lambda_\BreakL \BreakR \breakL
      + \frac{\breakQuad_\breakL v}{m_\BreakL} \breakR \BreakL
    \right)
    h
  + \text{h.c.}
\ .
\end{align}
For simplicity  we ignore this uncertainty in the current analysis by fixing the BRs to the values in relation~\eqref{eq:branching ratios}.

In order to be more consistent with the majority of the literature on collider phenomenology we will switch the notation from using Wely fermions \pair{T^c}{T} to Dirac fermions \pair{\overline T}{T} for the remaining part of this section.
We abbreviate heavy neutral bosons by $B^0 = \mathlist{Z,h}$ and assume that jets $j$ also include $b$-jets.
The decay channels in this notation are
\begin{subnumcases}{pp \to \overline \BreakL \BreakL h \to \label{eq:decay chains}}
    \bar t B^0 t B^0 h \to \bar t t j j (\bar b b) \ ,\label{eq:decay to tops}
 \\ \bar t B^0 b W^+ h \to \bar t j b j (\bar b b) \ , \label{eq:decay to top and bottom}
 \\ \bar b W^\pm b W^\mp h \to \bar b \bar l \nu b j (\bar b b) \ .\label{eq: decay to bottoms}
\end{subnumcases}
Here we assume that the associated Higgs is unboosted and decays into two separate bottom jets, while the other bosons are boosted and decay hadronically within the cone radius of a single jet.
Furthermore, we require the top quarks to decay either fully or semi-leptonically.

We have implemented Lagrangian~\eqref{eq:electroweak Lagrangian} together with the coupling which ensure the BRs~\eqref{eq:branching ratios} in \software[2.3]{FeynRules}~\cite{Alloul:2013bka}.
This enables us to generate events with \software[2.4]{MadGraph}~\cite{Alwall:2014hca} and decay heavy particles with \software{MadSpin}.
Light particle decay and showering is done in \software[6.4]{Pythia}~\cite{Sjostrand:2006za} followed by the detector simulation \software[3.3]{Delphes}~\cite{deFavereau:2013fsa}.

\begin{figure}
\begin{subfigure}{.5\textwidth}
\graphic{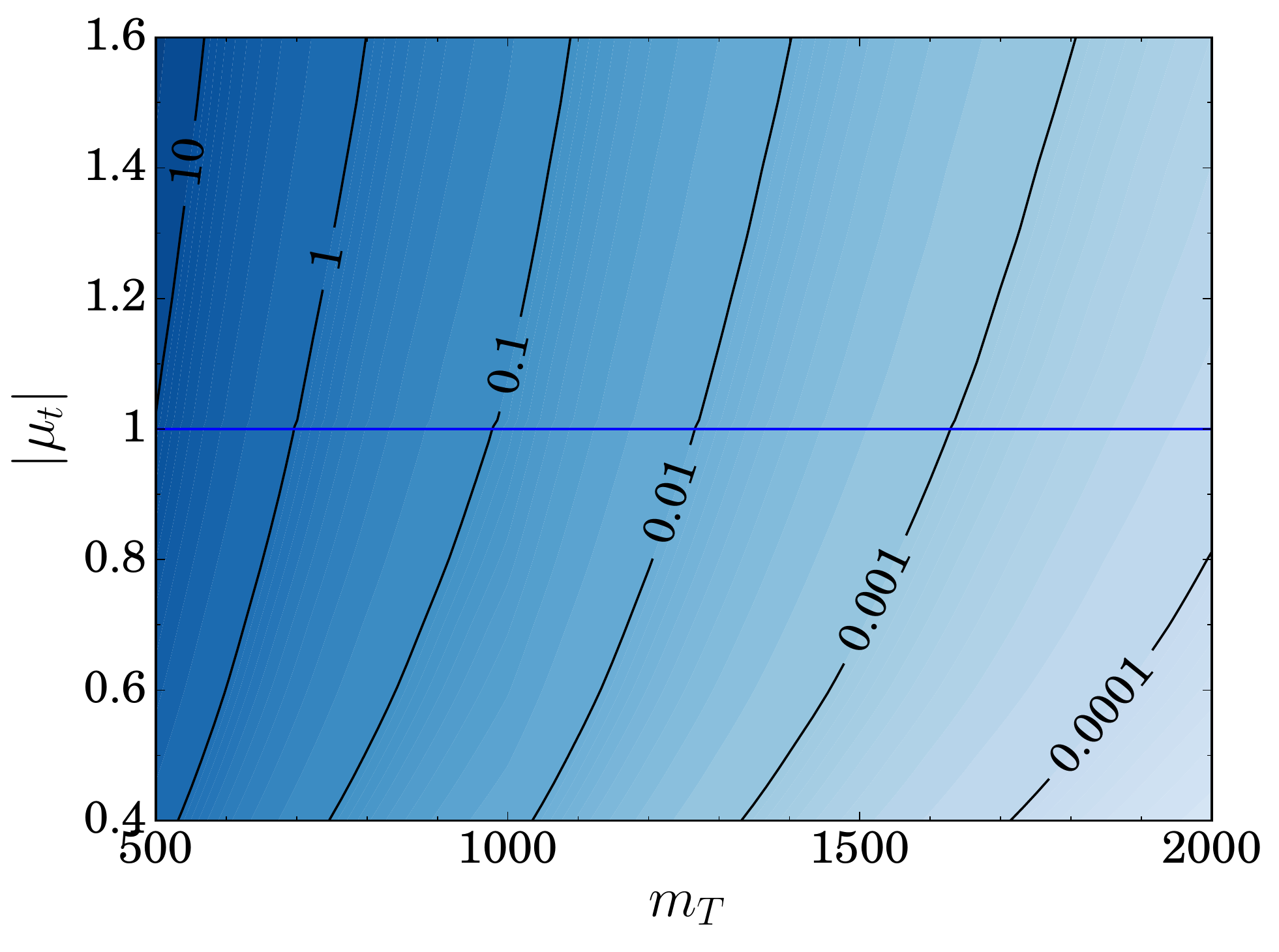}
\caption{\unit[14]{TeV}}
\label{fig:14TeV}
\end{subfigure}
\begin{subfigure}{.5\textwidth}
\graphic{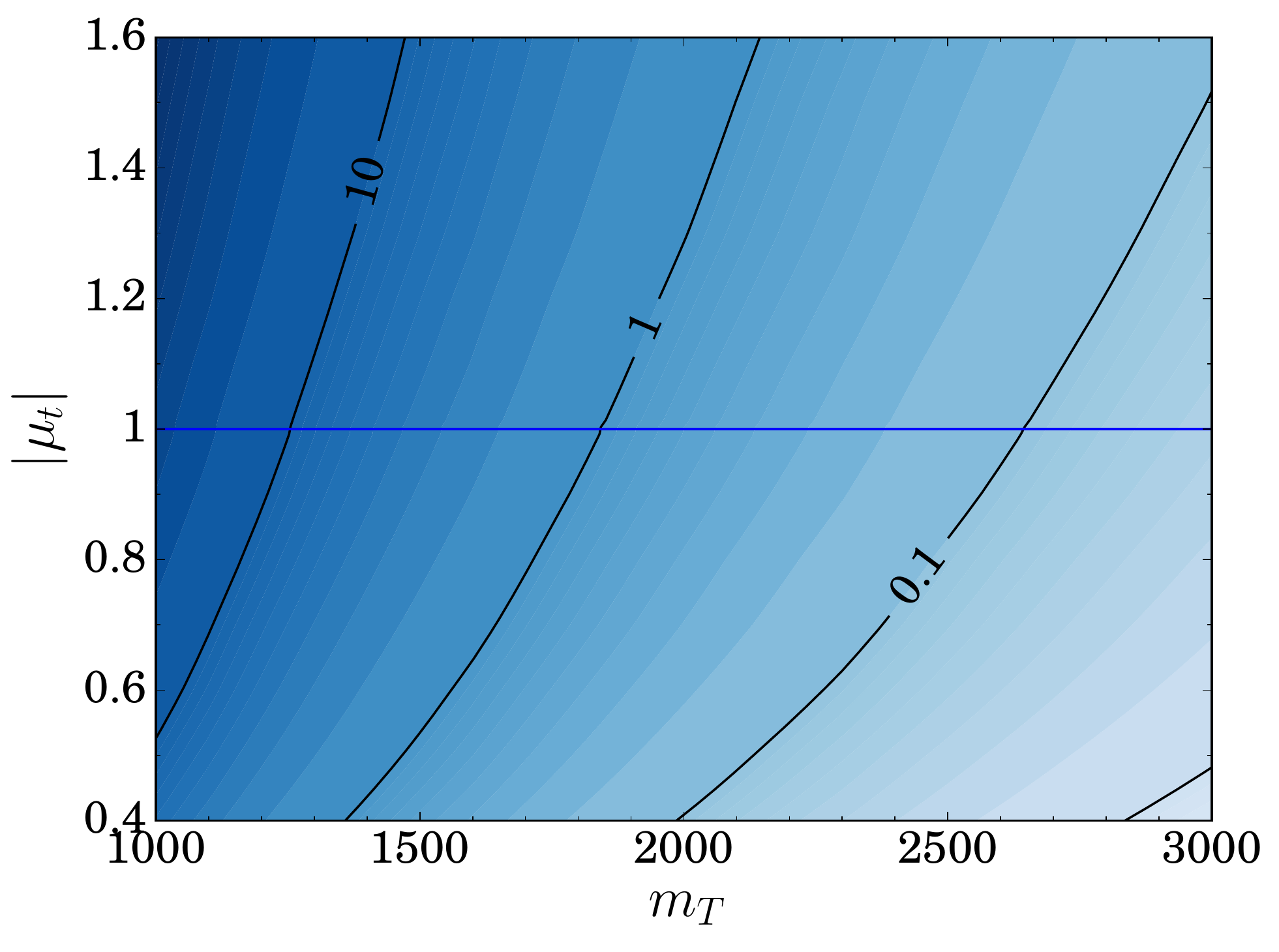}
\caption{\unit[100]{TeV}}
\label{fig:100TeV}
\end{subfigure}
\caption{%
Cross section of the signal process $\overline\BreakL\BreakL h$ at \unit[14]{TeV}~\subref{fig:14TeV} and \unit[100]{TeV}~\subref{fig:100TeV} in fb as a function of the top partner mass $m_\BreakL$ and the absolute value of the naturalness parameter $\abs{\mu_\breakL}$.
The horizontal line at $\abs{\mu_\breakL} = 1$ indicates natural models.
Note, that the production of $\overline TTh$ is not sensitive to the sign of $\mu_\breakL$, hence all figures depend on $\abs{\mu_\breakL}$ instead of $\mu_\breakL$.
} \label{fig:cross section}
\end{figure}

We have calculated the LO production cross section of the $\overline\BreakL\BreakL h$ process with \software{MadGraph}, and present it for \unit[14]{TeV} and \unit[100]{TeV} as a function of the top partner mass~$m_\BreakL$ and the naturalness parameter~$\mu_\breakL$ in Figure~\ref{fig:cross section}.
In particular, we neglect the Higgs coupling to two gluons, which is induced at one-loop level and formally a next-to-leading order contribution.
However, when $\abs{\mu_t}$ is small, the tree-level LO production rate becomes small and the one-loop contribution arising from the Higgs coupling to two gluons could become dominant.
For this reason we choose to focus on the region of parameter space where $\abs{\mu_t} > 0.4$.
We have checked that in this case the tree-level production rate still dominates.
For simplicity we have also assumed the top Yukawa coupling $\lambda_\breakL$ takes the value known from the SM, up to corrections of order $\nicefrac{v^2}{m_\BreakL^2}$.

\begin{table}
\centering
\begin{tabular}{lrr}
    \toprule
  & \multicolumn{2}{c}{Cross section [fb]}
 \\ \cmidrule{2-3}
  & semi-leptonic
  & full-leptonic
 \\ \midrule
    $p_T$
  & 3150
  & 526
 \\
    $n(l_\text{isol})$
  & 1315
  & 111
 \\ \bottomrule
\end{tabular}
\caption{%
Cross section of the background process $\bar ttjjjj$ at \unit[100]{TeV} in fb.
During event generation we require at least two quarks to be bottom or charm quarks with $p_T > \unit[40]{GeV}$.
The other two quarks are required to be boosted with $p_T > \unit[400]{GeV}$.
The resulting cross sections are given in the row labeled $p_T$.
Before the analysis we require the presence of at least one and two isolated leptons for semi- and full hadronic events, respectively.
The isolation criterion is given at the end of section~\ref{sec:simulation}.
We give the cross section after this pre-cut in the row labeled $n(l_\text{isol})$.
}
\label{tab:background cross section}
\end{table}

In order to reduce the hadronic background we focus on the decay channel~\eqref{eq:decay to tops} which involves two top quarks.
The dominant electroweak background for this signal consists of a pair of top quarks produced in association with three bosons.
In this case each top partner is mimicked by the combination of one top quark with one electroweak boson.
However, the top pair production in association with multiple quarks has a much larger cross section.
If we consider the mis-identification of jets we have to distinguish the cases in which pairs of jets mimic reconstructed objects from the cases in which a singe jet is mistaken for a single boosted object.
On the one hand, the decay products of heavy top-partner are generically boosted and tend to be detected as a single jet.
On the other hand, the Higgs is produced with a low transverse momentum and its decay products will be detected separately.
In this study we consider only Higgs decays into bottom quarks.
In order to reduce the size of the top pair plus jet background we require the two leading jets to be boosted enough to lie within a jet cone of $\Delta R \approx 0.5$
\begin{equation}
    p_T(j)
  > \unit[400]{GeV}
  \approx \frac{2 m_B}{\Delta R(j)}
\ .
\end{equation}
We expect these jets to mimic the bosons originating from the top partner decay.
Additionally, a combination of two softer jets can be misidentified as being the Higgs boson.
As we require the Higgs boson signature to be $b$-like we restrict these jets to come from $c$ or $b$ quarks.
Therefore, the main background in this analysis is top pair production in association with four jets where the two leading once are boosted while the other two are $b$-like.
We have generated this background with \software{MadGraph} and have required the jet transverse momentum to be $p_T > \unit[400\text{ and }40]{GeV}$ for the two leading and the remaining jets, respectively.
We show the LO cross sections for semi- and full leptonic decays of the top quarks in Table~\ref{tab:background cross section}.
As a cut on the jet number is only a weak discriminator the top-pair production with more than four jets will contribute to the background.
Because the strong coupling constant appears in an higher power and due to the increased number of jets which leads to a larger phase space this higher order background will be suppressed with respect to the leading contribution.
We do not consider any $k$-factor in this analysis.

During the analysis we cluster jets with the anti-$k_T$ algorithm using a minimal transverse momentum of \unit[40]{GeV} and a jet cone size of 0.5.
Additionally we require the two leading jets to be boosted more than $p_T > \unit[400]{GeV}$.
We require at least one and two leptons for the semi- and full-leptonic analysis, respectively.
Leptons must be isolated with an isolation radius of $\Delta R > 0.3$ and a maximal transverse momentum ratio of 0.2.
The transverse momentum ratio is defined between the lepton and other cell activity within the isolation cone.
If the lepton transverse momentum is larger than \unit[100]{GeV}, we do not require it to be isolated~\cite{Brust:2014gia}.
Finally, we disregard events with less than five jets.

\subsection{Event reconstruction}

\begin{figure}
\begin{subfigure}{.5\textwidth}
\graphic{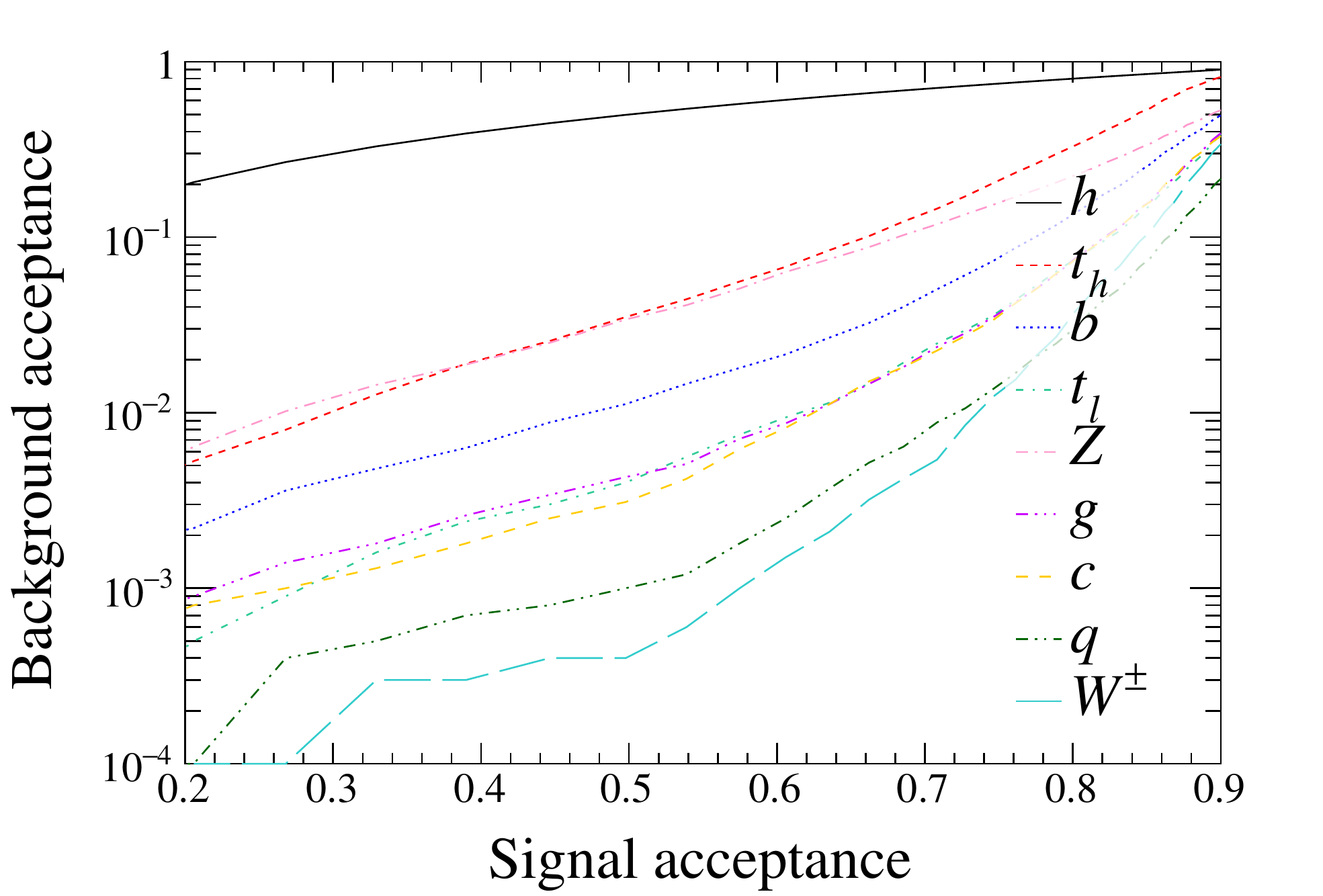}
\caption{Higgs tagger for $\unit[400]{GeV} < p_T < \unit[700]{GeV}$.}
\label{fig:higgs ROC low}
\end{subfigure}
\begin{subfigure}{.5\textwidth}
\graphic{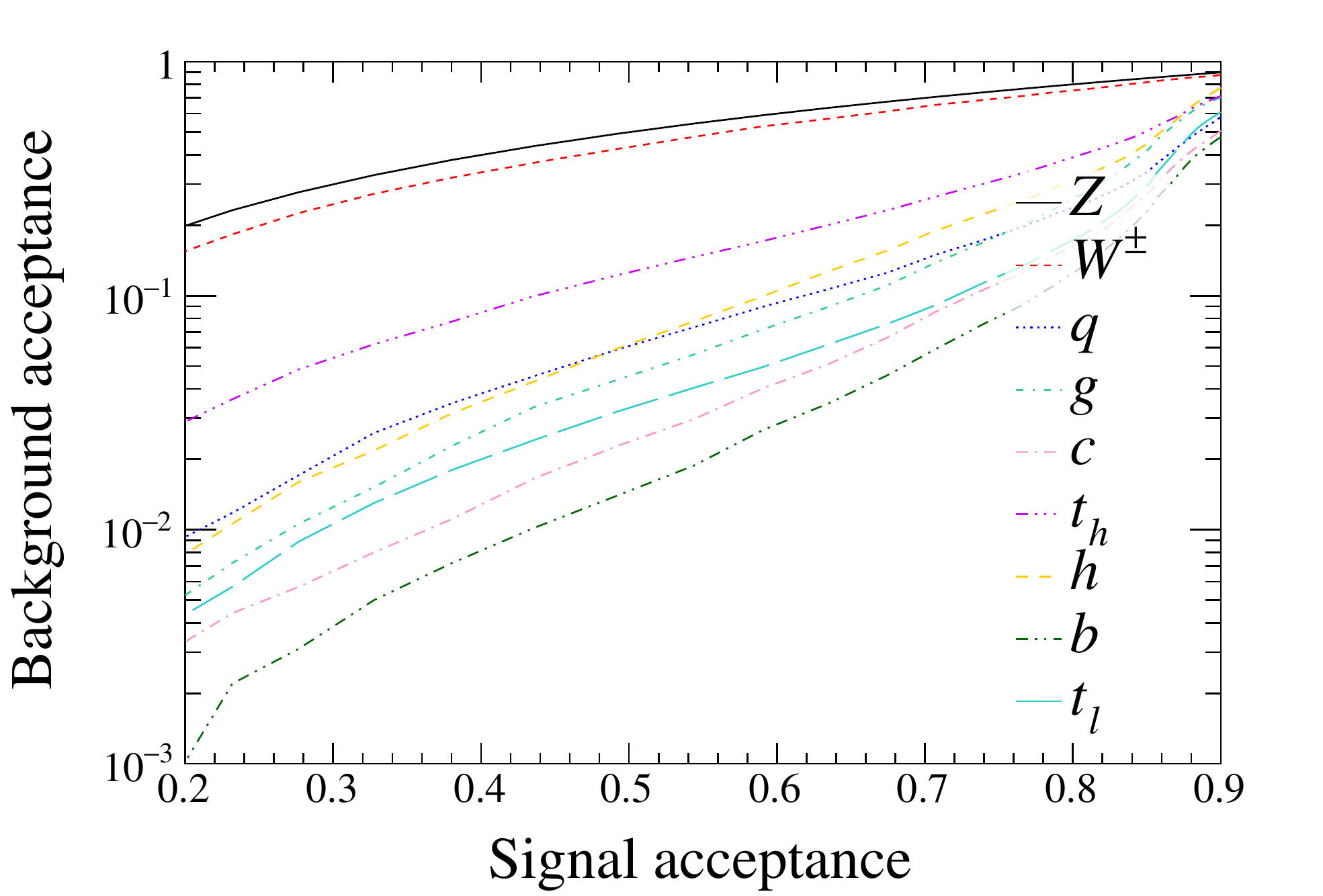}
\caption{$Z$ tagger for $\unit[400]{GeV} < p_T < \unit[700]{GeV}$.}
\label{fig:Z ROC low}
\end{subfigure}
\begin{subfigure}{.5\textwidth}
\graphic{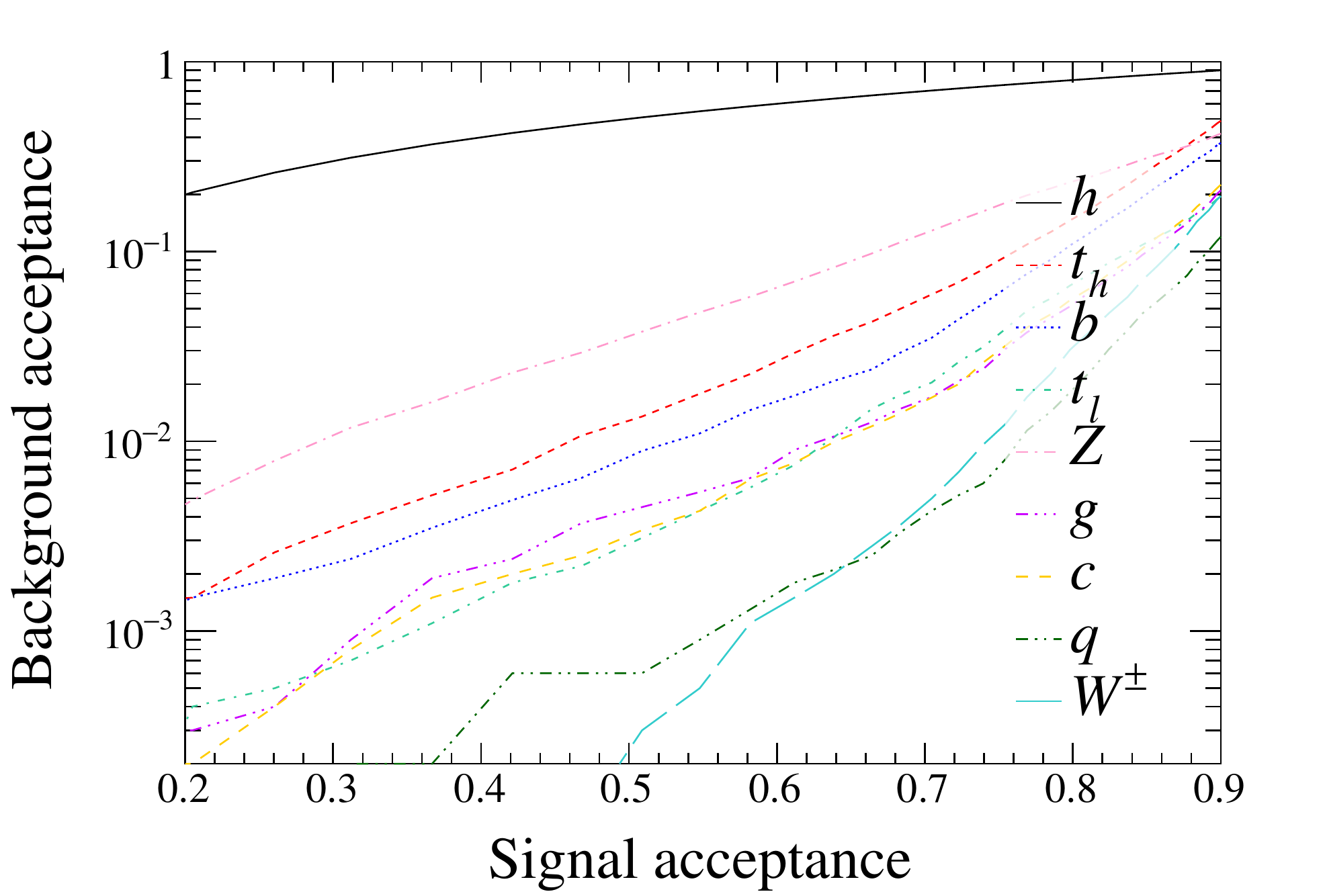}
\caption{Higgs tagger for $\unit[700]{GeV} < p_T < \unit[1000]{GeV}$.}
\label{fig:higgs ROC high}
\end{subfigure}
\begin{subfigure}{.5\textwidth}
\graphic{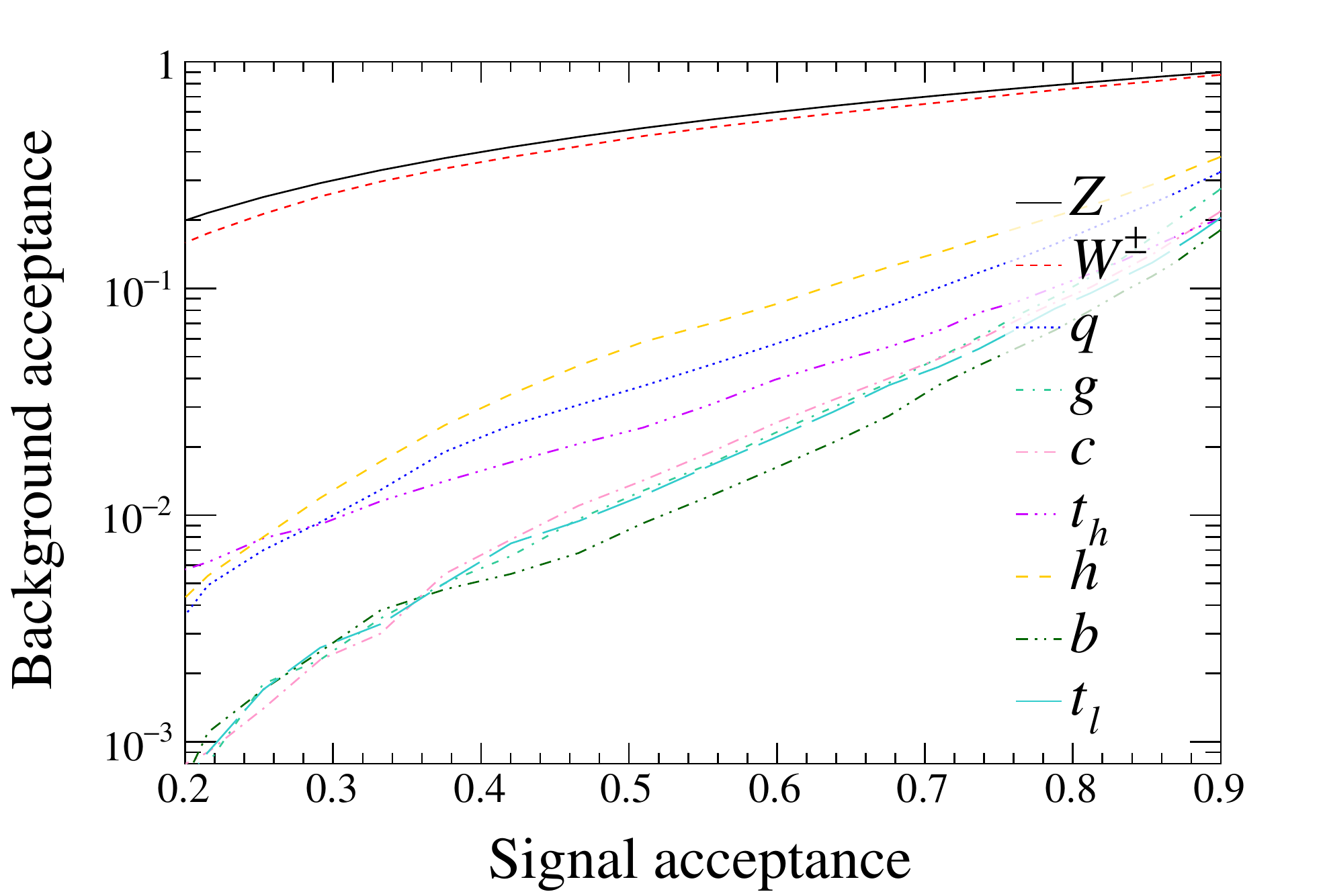}
\caption{$Z$ tagger for $\unit[700]{GeV} < p_T < \unit[1000]{GeV}$.}
\label{fig:Z ROC high}
\end{subfigure}
\caption{%
ROC curves of the signal and background acceptance for the Higgs tagger in~\subref{fig:higgs ROC low} and~\subref{fig:higgs ROC high} as well as for the $Z$-boson tagger in~\subref{fig:Z ROC low} and~\subref{fig:Z ROC high}.
All processes are generated at \unit[100]{TeV} via pair-production.
(We are using an effective model in order to produce Higgs boson pairs.)
During production and tagging we have ensured that the transverse momenta are constrained to \unit[400--700]{GeV} for~\subref{fig:higgs ROC low} and~\subref{fig:Z ROC low} and to \unit[700--1000]{GeV} for~\subref{fig:higgs ROC high} and~\subref{fig:Z ROC high}.
As the jet cone size is $\Delta R = 0.5$, we expect the bosons of the SM and the top quarks to lie within a single jet for $p_T \gtrsim \unit[400\text{ and }700]{GeV}$, respectively.
}
\label{fig:boson tagger}
\end{figure}

We reconstruct the signal signatures with multiple boosted decision trees (BDTs).
Each of them is optimized to tag or reconstruct a single physical object, such as bottom or top quark, Higgs or $Z$ boson and top partners.
Such an object can consist of a sub-jet, a single jet or be combined from multiple jets.
This Boosted Collider Analysis (\software{BoCA})~0.3~\cite{Boca} has been introduced in~\cite{Hajer:2015gka, Craig:2016ygr} where we also discuss the efficiency of the bottom and top taggers.
\software{BoCA} uses the \software{TMVA} library~\cite{Hocker:2007ht} of the \software[6.06]{ROOT} framework~\cite{Brun:1997pa} together with \software[3.2]{FastJet}~\cite{Cacciari:2011ma} for jet clustering.
Here we introduce additionally taggers dedicated to Higgs and $Z$ boson tagging.
For boosted bosons we demonstrate their background mis-identification as a function of a given signal rate, the so called Receiver Operating Characteristic (ROC) curves in Figure~\ref{fig:boson tagger}.
We show the results for two different ranges of transverse momentum, which correspond to the cases in which the top quarks are boosted enough to end up in one and two jets, respectively.
The Higgs tagger utilizes the displaced vertices of the bottom decay products.
The strongest discriminators within these taggers are the boson mass and the pull~\cite{Gallicchio:2010sw} between its components.
Furthermore, we train a neutral boson tagger to detect both of these particles, which allows us to reconstruct top partners from tops and neutral bosons.
During this reconstruction we cater towards the boostness of the decay products by using the available substructure information.
In order to suppress the $\bar tt+3j$ and the $\bar tt + 5j$ backgrounds we require the neutral bosons from the top partner decays to be boosted with a cone size of $\Delta R < 0.5$ while the associated Higgs boson has to be un-boosted with a cone size of $\Delta R > 0.5$.
Finally, we reconstruct the $\overline TTh$ signature by combining two top partner tagger with a Higgs boson tagger.
The signal is separated from the background with a single cut on the final BDT result.

\subsection{Results}\label{sec:results}

\begin{figure}
\begin{subfigure}{.5\textwidth}
\graphic{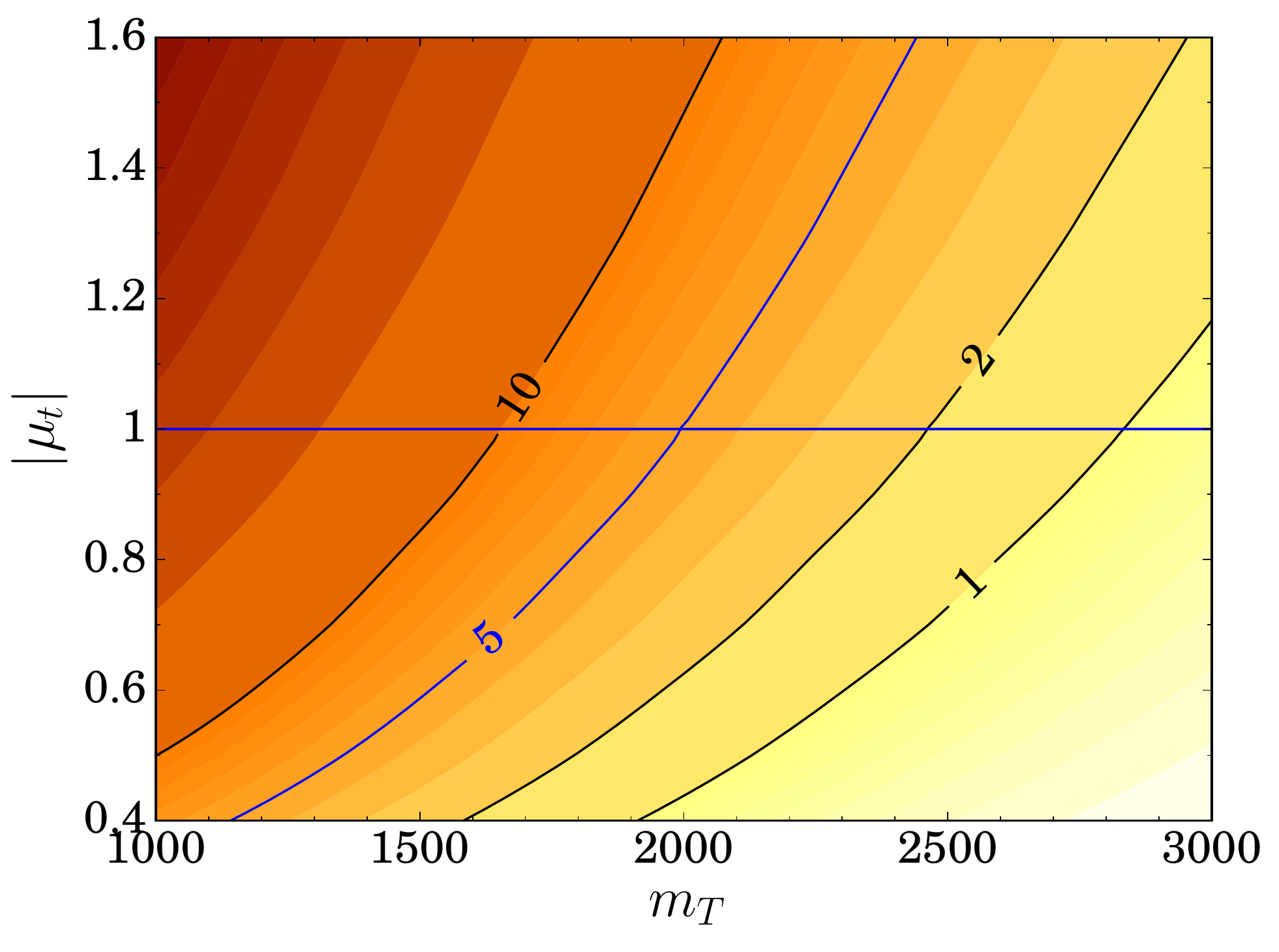}
\caption{Luminosity of $\unit[3]{ab^{-1}}$.}
\label{fig:discovery fixed lumi}
\end{subfigure}
\begin{subfigure}{.5\textwidth}
\graphic{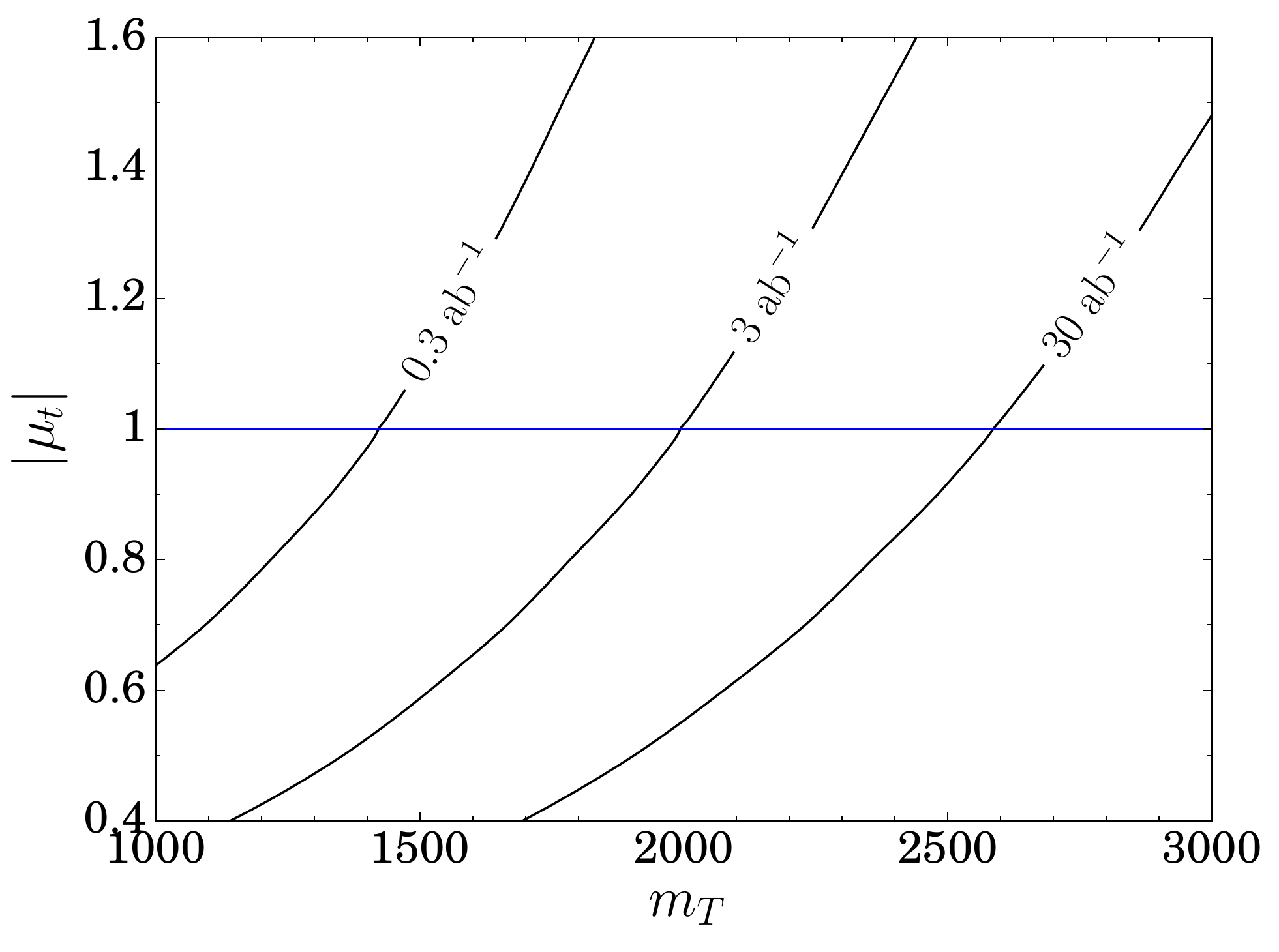}
\caption{Significance of 5$\sigma$.}
\label{fig:discovery fixed significance}
\end{subfigure}
\caption{%
Discovery reach defined as $\sig{b}{b + s}$ for pair production of top partners in association with one Higgs boson at \unit[100]{TeV}.
We present the reaches for a fixed luminosity of $\unit[3]{ab^{-1}}$ in Figure~\subref{fig:discovery fixed lumi} and for a fixed significance of 5$\sigma$ with luminosities of $\listunit[0.3, 3, 30]{ab^{-1}}$ in Figure~\subref{fig:discovery fixed significance}.
}
\label{fig:discovery}
\end{figure}

For $n$ observed events the significance can be defined as the log-likelihood ratio~\cite{Cowan:2010js}
\begin{equation}
    \sig{x}{n}
  = \sqrt{- 2 \ln \frac{\like{x}{n}}{\like{n}{n}}}
\ , \label{eq:significance}
\end{equation}
where $x$ is the number of events predicted by the hypothesis which is tested.
For counting experiments the likelihood function is given by the Poisson probability
\begin{equation}
    \like{x}{n}
  = \frac{x^n e^{-x}}{n!}
\ .
\end{equation}
For the exclusion of the signal hypothesis $b + s$ we require $\sig{b + s}{n} \geq 2$ and for the discovery of a signal by exclusion of the background only hypothesis $b$ we require $\sig{b}{n} \geq 5$.
As we are projecting to future experiments we replace the event number $n$ with the prediction for an alternative hypothesis.
Hence, we are using $\sig{b}{b + s} \geq 5$ for the discovery of a model and $\sig{b + s}{\at{b + s}{\text{nat}}} \geq 2$ for the exclusion of an unnatural model.

\begin{figure}
\centering
\graphic[0.5]{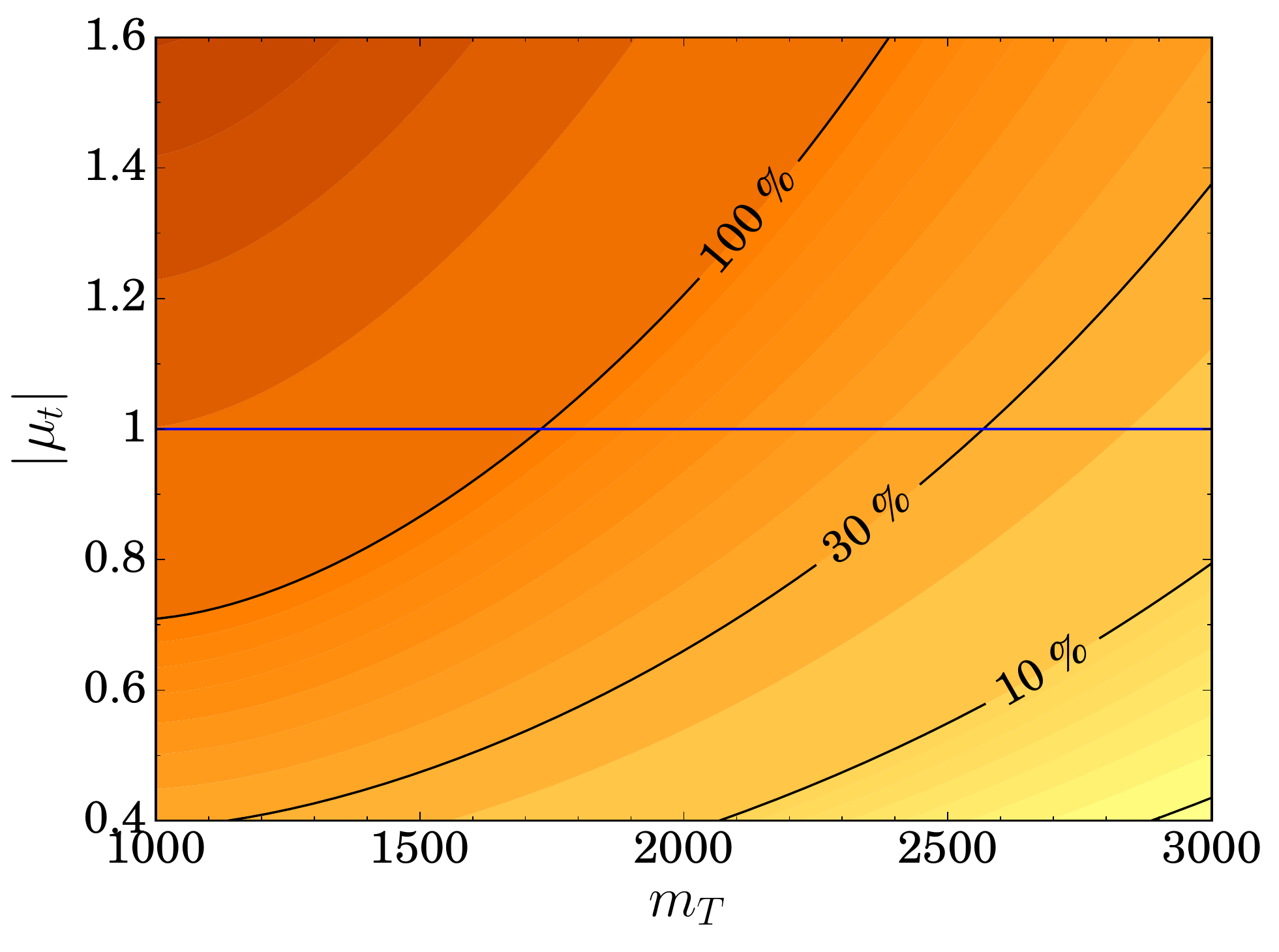}
\caption{Ratio of signal over background $\nicefrac{s}{b}$.}
\label{fig:s over b}
\end{figure}

In Figure~\ref{fig:discovery}, we present the discovery reach against the background-only hypothesis of top partners in the $\overline\BreakL\BreakL h$ production channel at \unit[100]{TeV}.
This may not be the most sensitive channel for the search of fermionic top partners, but it shows the effectiveness of our analysis for testing naturalness.
For natural theories we see that the discovery reach for top partners coupled to a Higgs boson can be pushed up to $\sim\listunit[1.4, 2, 2.5]{TeV}$, with $\listunit[0.3, 3, 30]{ab^{-1}}$ of data, respectively.
In comparison, the ATLAS~\cite{1434358, Aaboud:2016lwz, ATLAS:2016sno} and CMS~\cite{CMS:2016ete} experiments have excluded fermionic top partner with $m_\BreakL < \unit[780]{GeV}$, given BRs compatible with~\eqref{eq:branching ratios}.
The systematic errors are not included in our analysis.
Instead, we present the expected signal over background ratio in Figure~\ref{fig:s over b} which indicates that the contour with $\nicefrac{s}{b} = \unit[10]{\%}$ could extend to above $\unit[2.5]{TeV}$ for natural theories.

\begin{figure}
\begin{subfigure}{.5\textwidth}
\graphic{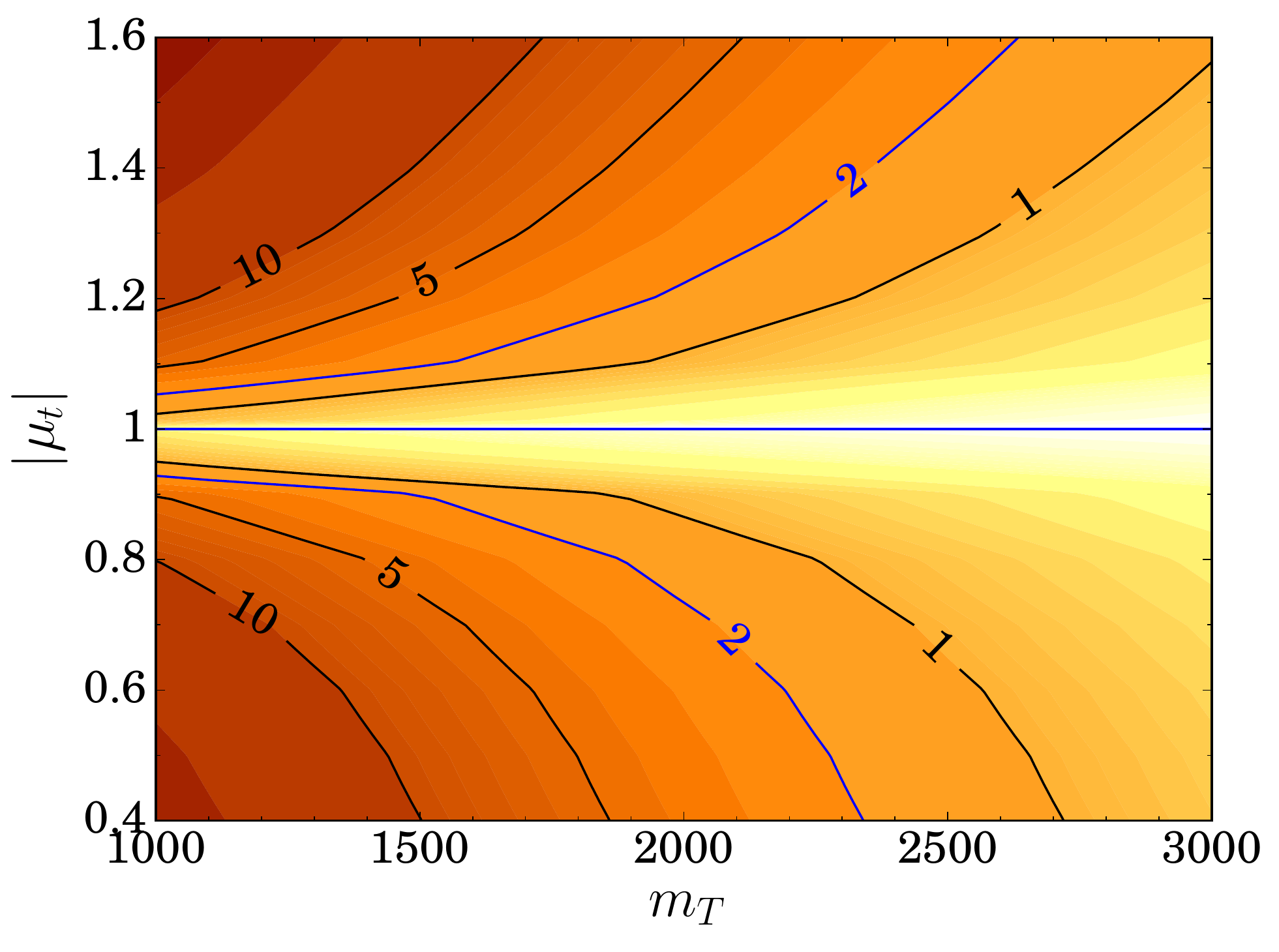}
\caption{Luminosity of $\unit[3]{ab^{-1}}$.}
\label{fig:exclusion fixed lumi}
\end{subfigure}
\begin{subfigure}{.5\textwidth}
\graphic{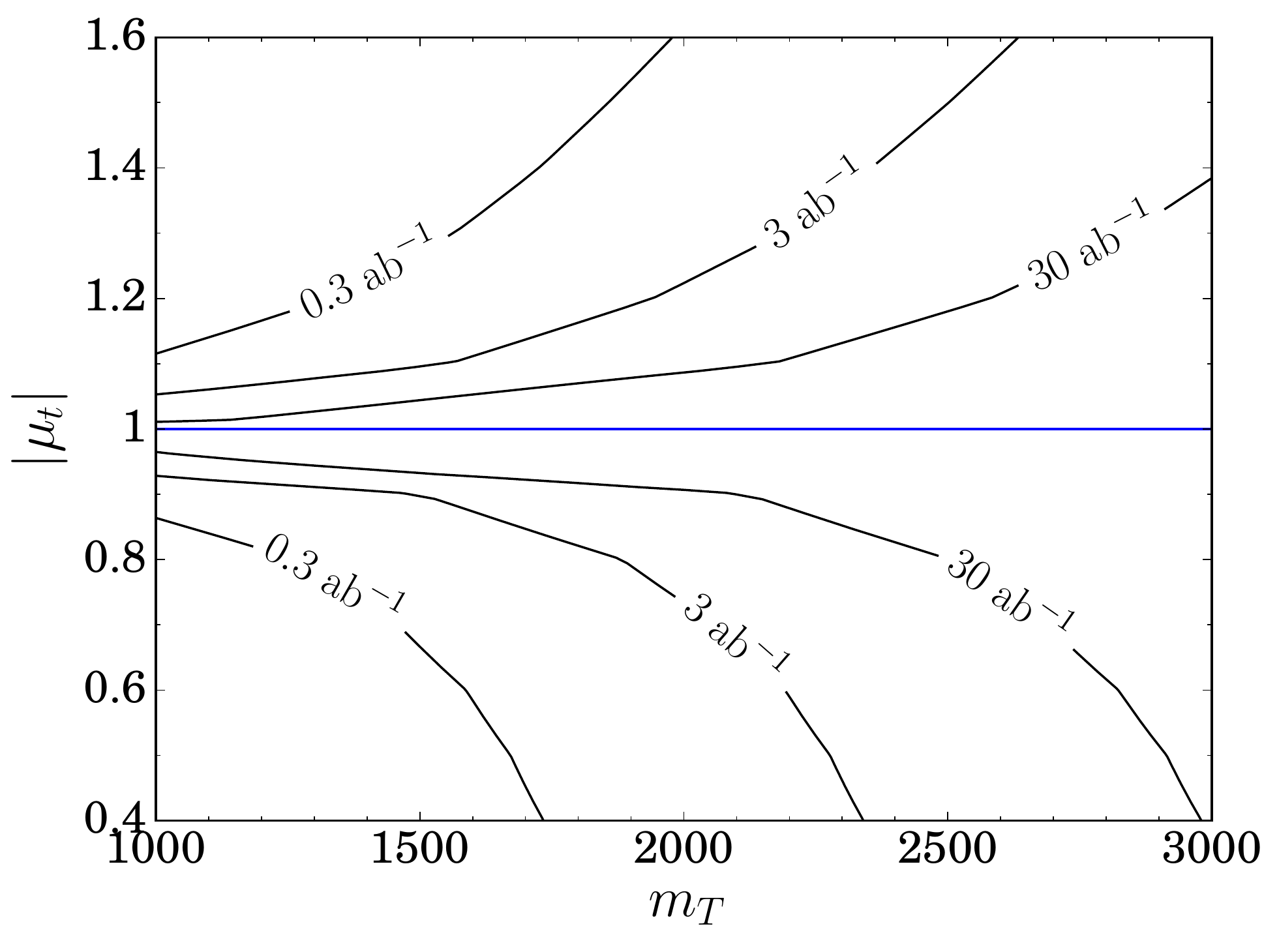}
\caption{Significance of 2$\sigma$.}
\label{fig:exclusion fixed significance}
\end{subfigure}
\caption{%
Exclusion limit for unnatural theories defined by $\sig{b + s}{\at{b + s}{\text{nat}}}$ as a function of the top partner mass.
Based on $\overline\BreakL\BreakL h$ production at \unit[100]{TeV}.
For a luminosity of $\unit[3]{ab^{-1}}$ in Figure~\subref{fig:exclusion fixed lumi} and for $\listunit[0.3, 3, 30]{ab^{-1}}$ and a fixed significance of 2$\sigma$ in Figure~\subref{fig:exclusion fixed significance}.
}
\label{fig:exclusion}
\end{figure}

In the following we introduce two measures for the sensitivity of testing the naturalness sum rule.
First, we calculate the exclusion limit of unnatural theories with $\abs{\mu_\breakL} \neq 1$, against the assumption that the observation at the collider is consistent with the prediction of a theory with $\abs{\mu_\breakL} = 1$.
The exclusion limits of unnatural theories are presented in Figure~\ref{fig:exclusion}.
With $\unit[30]{ab^{-1}}$ of data, unnatural theories with $\abs{\abs{\mu_\breakL} - 1} > 0.1$ can be excluded up to $m_\BreakL \sim \unit[2.2]{TeV}$.

\begin{figure}
\begin{subfigure}{.5\textwidth}
\graphic{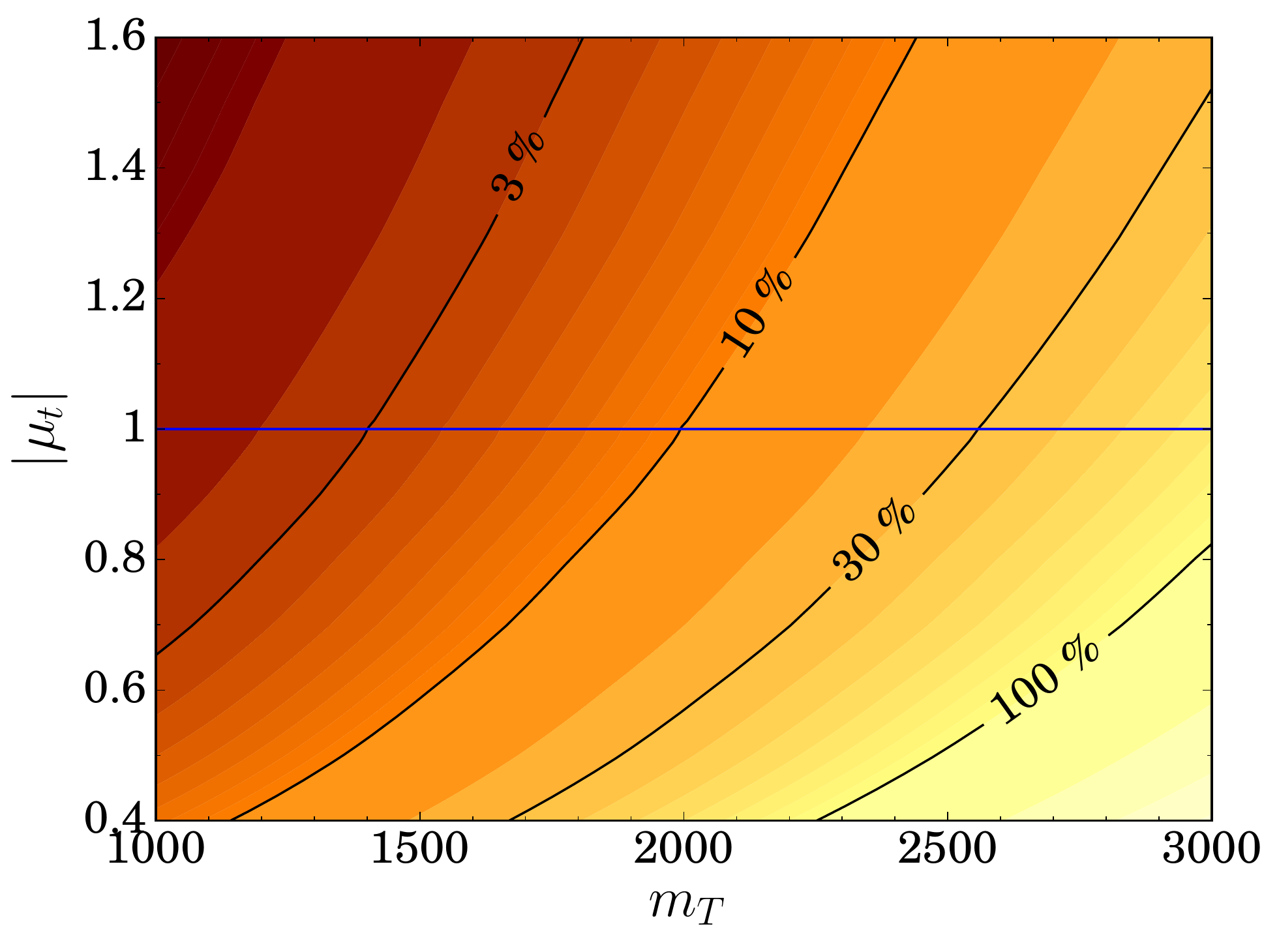}
\caption{Luminosity of $\unit[3]{ab^{-1}}$.}
\label{fig:precision fixed lumi}
\end{subfigure}
\begin{subfigure}{.5\textwidth}
\graphic{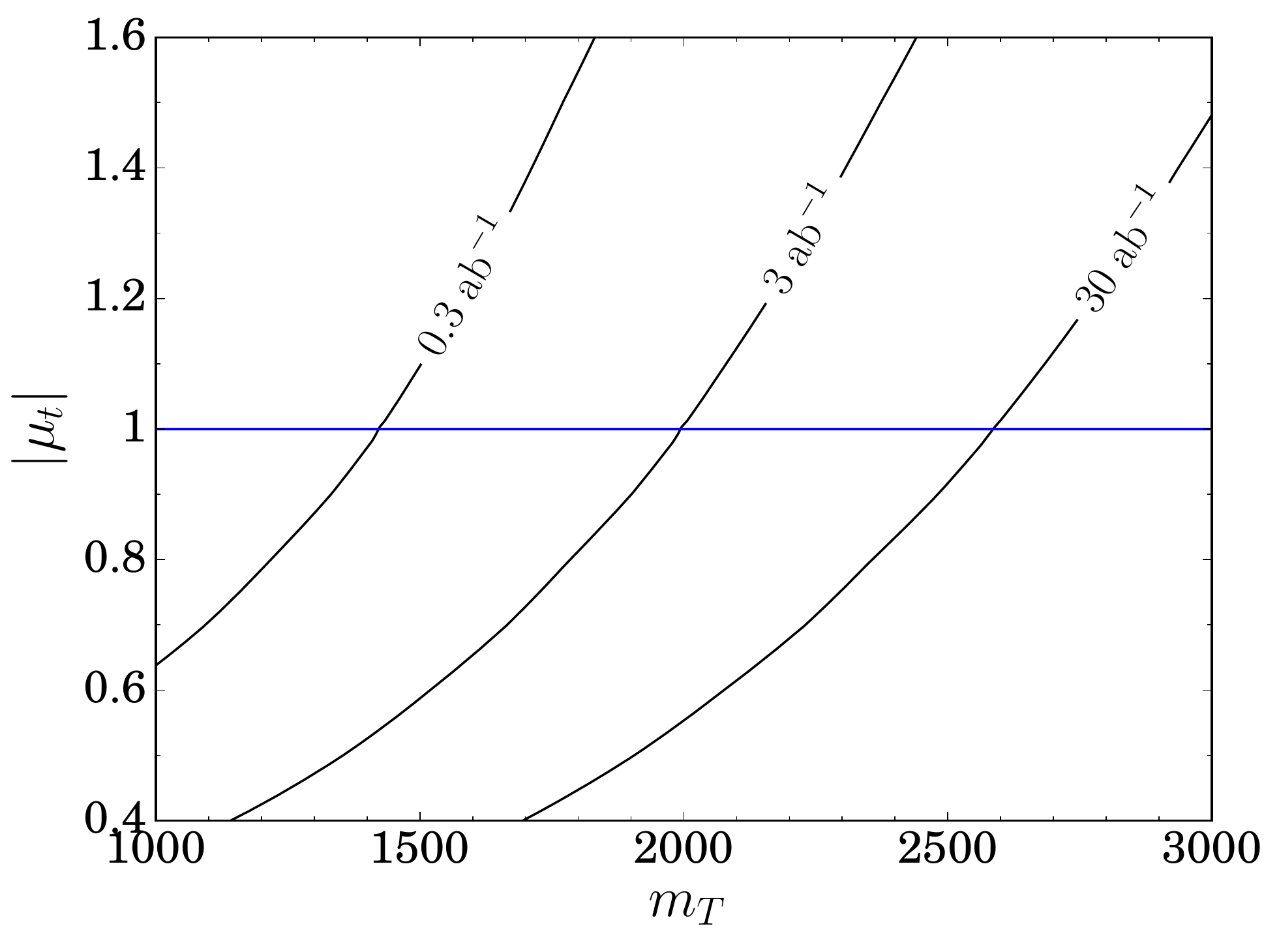}
\caption{Precision of \unit[10]{\%}.}
\label{fig:precision fixed percentage}
\end{subfigure}
\caption{%
Precision of measuring the absolute value of the naturalness parameter $\abs{\mu_\breakL}$ at $\unit[3]{ab^{-1}}$ in Figure~\subref{fig:precision fixed lumi}, and the reach of measuring with a precision of \unit[10]{\%} at $\listunit[0.3, 3, 30]{ab^{-1}}$ in Figure~\subref{fig:precision fixed percentage}.
Here we assume that the top Yukawa coupling can be perfectly measured.
}
\label{fig:precision}
\end{figure}

Second, we calculate the precision of measuring the naturalness parameter, defined as the uncertainty at one sigma confidence level
\begin{equation}
    \delta \abs{\mu_\breakL}
  = \sqrt{
      \left(- \frac{1}{\lambda_\breakL^2} \delta \breakQuad_\BreakL\right)^2
    + \left(2 \frac{\breakQuad_\BreakL}{\lambda_\breakL^3} \delta \lambda_\breakL\right)^2
    }
\ . \label{eq:precision}
\end{equation}
The precision of the coupling $\breakQuad_\BreakL$ can be estimated with the relation
\begin{equation}
    \delta \breakQuad_\BreakL
  = \frac{\breakQuad_\BreakL}{2 \sig{b}{n}}
\ .
\end{equation}
In Figure~\ref{fig:precision} we present the precision under the assumption of a perfectly measured top Yukawa coupling $\delta \lambda_\breakL = 0$.
Figure~\ref{fig:precision fixed lumi} shows that in natural theories the naturalness parameter could be measured with a precision of \unit[10]{\%} up to a top partner mass of $m_\BreakL \sim \unit[2]{TeV}$, using $\unit[3]{ab^{-1}}$ of data.
Figure~\ref{fig:precision fixed percentage} shows that with $\listunit[0.3, 3, 30]{ab^{-1}}$ of data a precision of \unit[10]{\%} could be achieved for top partners in natural theories with masses up to $\sim\unit[1.4]{TeV}$, $\sim\unit[2]{TeV}$ and above \unit[2.5]{TeV}, respectively.
Note that the definition of the precision~\eqref{eq:precision} has the consequence that curves for $\sig{b}{n} = 5$ presented in Figure~\ref{fig:discovery} are approximately identical to the curves for $\nicefrac{\delta \abs{\mu_\breakL}}{\abs{\mu_\breakL}} = \unit[10]{\%}$ shown in Figure~\ref{fig:precision}.

\begin{figure}
\begin{subfigure}{.5\textwidth}
\graphic{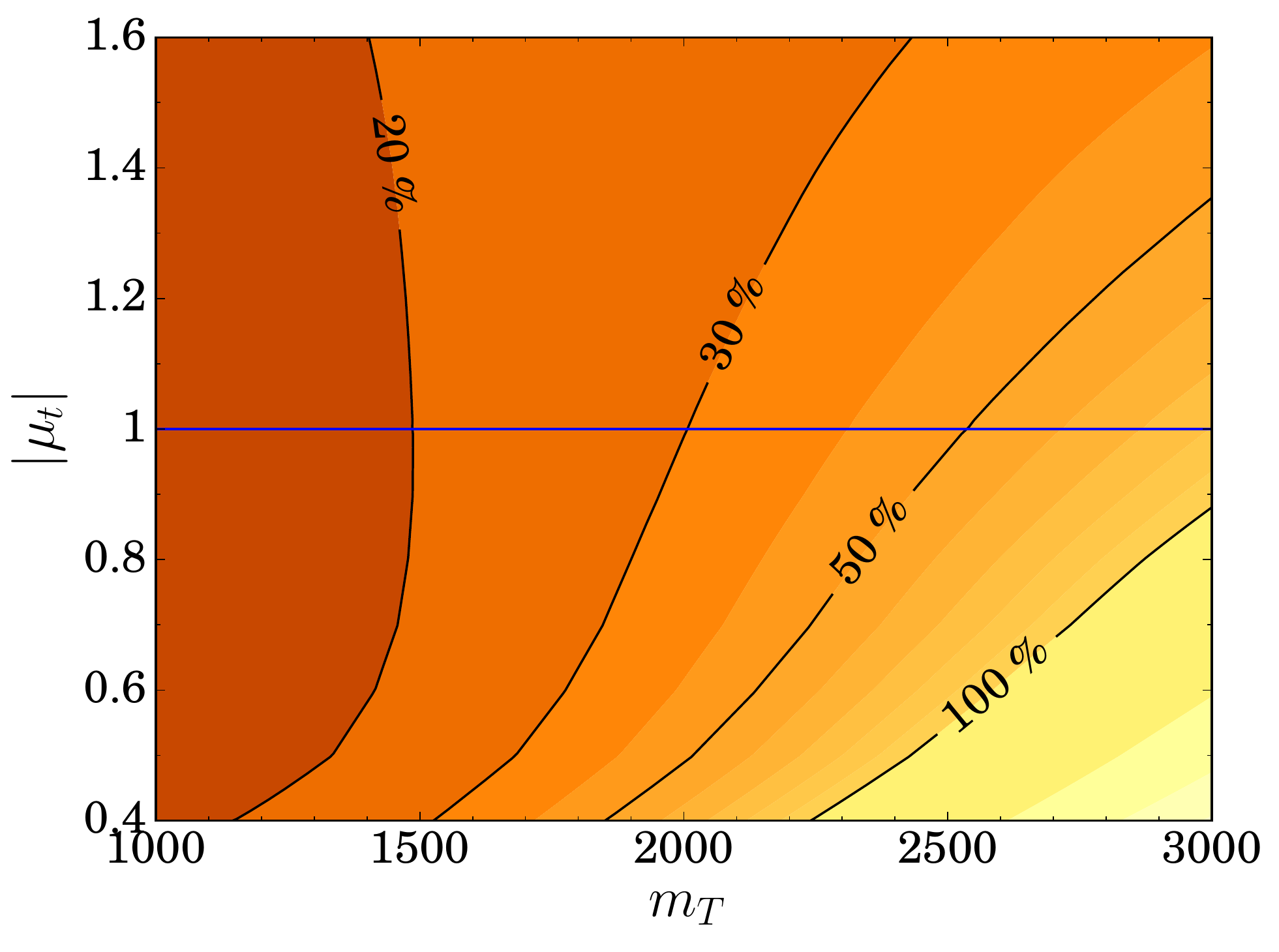}
\caption{Luminosity of $\unit[3]{ab^{-1}}$ and $\delta \lambda_\breakL = \unit[10]{\%}$.}
\label{fig:precision lumi 3ab}
\end{subfigure}
\begin{subfigure}{.5\textwidth}
\graphic{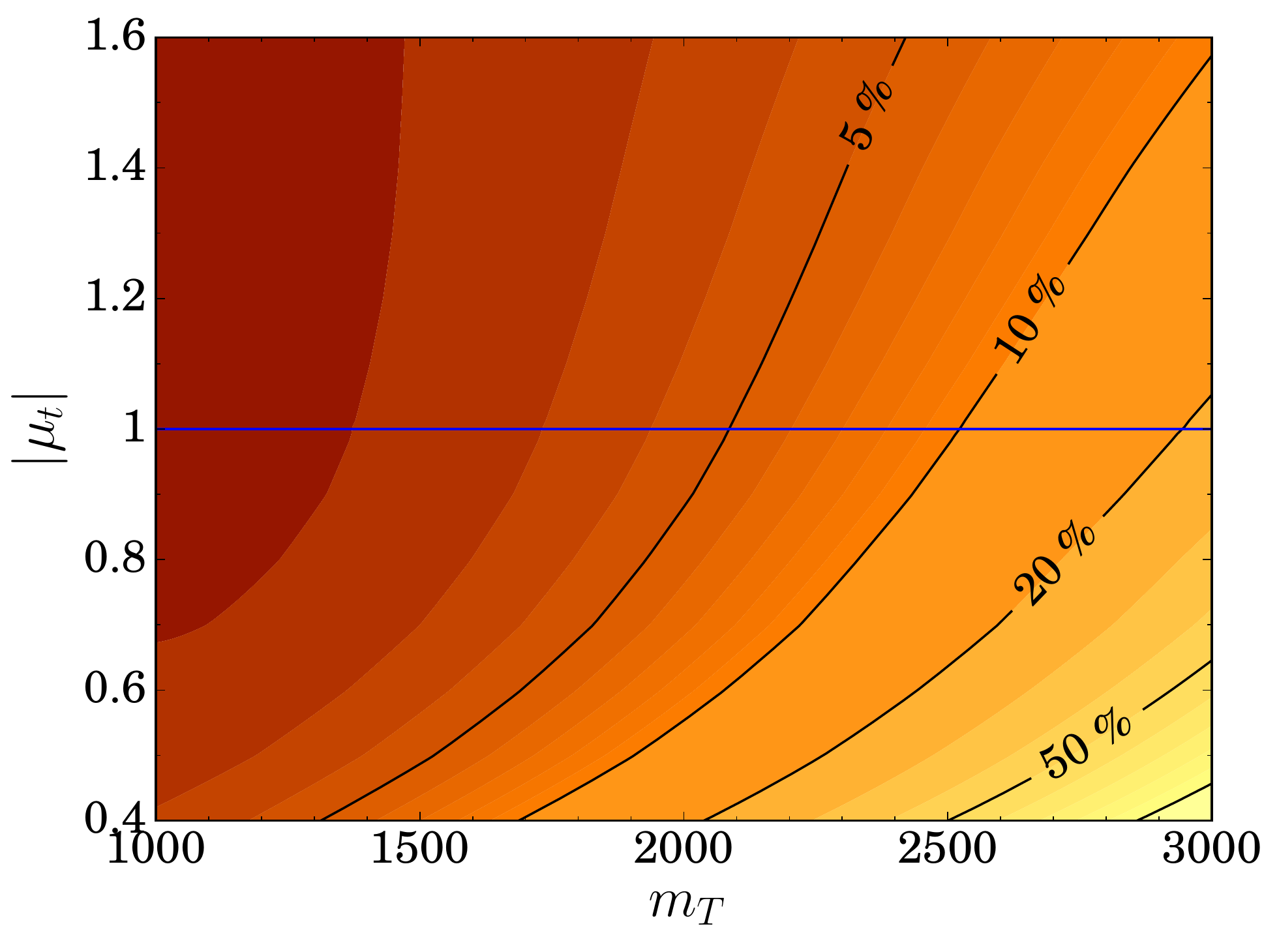}
\caption{Luminosity of $\unit[30]{ab^{-1}}$ and $\delta \lambda_\breakL = \unit[1]{\%}$.}
\label{fig:precision lumi 30ab}
\end{subfigure}
\caption{%
Precision of measuring the absolute value of the naturalness parameter $\abs{\mu_\breakL}$ at \unit[100]{TeV} with $\unit[3\text{ and }30]{ab^{-1}}$ of data.
We assume the uncertainty of measuring $\lambda_\breakL$ to be \unit[10]{\%} and \unit[1]{\%}, respectively.
} \label{fig:precision 2}
\end{figure}

In Figure~\ref{fig:precision} we have neglected the uncertainty on the measurement of the top Yukawa coupling.
However, the measurement of the top Yukawa coupling is non-trivial and will be an important task at a \unit[100]{TeV} hadron collider~\cite{Plehn:2015cta}.
In Figure~\ref{fig:precision 2}, we show the precision of measuring the naturalness parameter when the uncertainty of measuring $\lambda_\breakL$ is included.
We consider two cases.
In Figure~\ref{fig:precision lumi 3ab} we assume a luminosity of $\unit[3]{ab^{-1}}$ at \unit[100]{TeV} and an uncertainty of \unit[10]{\%} achievable in future $\lambda_\breakL$ measurements at the HL-LHC~\cite{CMS-NOTE-2012-006, ATLAS-collaboration:1484890}.
In Figure~\ref{fig:precision lumi 30ab} we consider $\unit[30]{ab^{-1}}$ and an uncertainty of \unit[1]{\%} potentially reachable at a future \unit[100]{TeV} collider~\cite{Plehn:2015cta}.
Although a precision of \unit[10]{\%} for the measurement of the naturalness parameter could still be pushed up to $m_\BreakL \sim \unit[2.5]{TeV}$, we do see a significant impact from $\delta \lambda_\breakL$, particularly for small $m_\BreakL$.
Therefore, it is crucial to improve the precision of measuring $\lambda_\breakL$, in order to test naturalness with a higher precision.

\section{Discussion and Conclusion}\label{sec:conclusion}

The naturalness problem plays an extremely important role in driving particle physics.
While, searches for partner particles predicted by theories of naturalness are actively conducted at the LHC, a further step is necessary to actually test the naturalness of these particles.
Once discovered, it must be ensured that the newly found particle is not an impostor, but a particle with the right properties to cancel the quadratic UV-sensitivity of the Higgs self-energy.

Using the fermionic top sector as an example, we show how the naturalness sum rule can be measured and what sensitivity reach can be achieved at a \unit[100]{TeV} hadron collider.
We developed a simplified model for vector-like fermionic top partners and derived the naturalness sum rule in this general setup.
Remarkably, we found that the naturalness sum rule only involves couplings of the Higgs to a pair of top quarks and a pair of top partners, and is exact up to an order of $\nicefrac{v^2}{m_\BreakL^2}$.
The derived condition is rather generic, covering representative theories of naturalness, including for example little Higgs models both with and without $T$-parity and mirror twin Higgs models.
In particular, the necessary measurements do not involve the measurement of the decay constant~$f$.
The \unit[100]{TeV} collider analyses based on the $\overline\BreakL\BreakL h$ production in the little Higgs models indicates that at $\unit[30]{ab^{-1}}$:
\begin{itemize}
\item Theories which predict a $\abs{\mu_\breakL}$ deviating from the naturalness assumption by more than \unit[10]{\%} which corresponds to $\abs{\abs{\mu_\breakL} - 1} > 0.1$ could be excluded for a top partner mass of up to $\sim\unit[2.2]{TeV}$.
\item A precision of \unit[10]{\%} for the measurement of $\abs{\mu_\breakL}$ could be achieved with a top partner mass of up to $\sim\unit[2.5]{TeV}$, given the precision of measuring the top Yukawa coupling of some percent.
\end{itemize}

We expect that the reach of this measurement can be further improved.
Recall, that in this analysis presented in this article we only consider decays of $\overline TTh$ via a pair of top quarks and require them to decay fully or semi-leptonically.
However, a larger branching fraction originates from the decay via one top quark and one bottom quark, that is $\overline TTh \to \bar t B^0 b W^+ h$ or $\overline TTh \to t B^0 \bar b W^- h$.
Additionally, the Higgs boson accompanying the top partners are required to  decay to a pair of bottom quarks.
In reality, some of its other decay modes may lead to a stronger background suppression.
We would expect that these channels bring non-trivial improvements to the sensitivity of testing naturalness.
Despite its important role in testing naturalness,  we would like to stress again that the $\overline TTh$ production is insensitive to the sign of the naturalness parameter and hence causes a degeneracy problem.
In order to break this degeneracy and exclude the theories with $\mu_\breakL < 0$, new strategies such as indirect detection needs to be introduced.

Given the crucial role of the naturalness problem in driving particle physics, it is worthwhile to extend the analysis pursued in this article to other contexts.
On the one hand, a complete cancellation of quadratic UV-sensitivity in the Higgs self-energy in little Higgs models requires the introduction of additional partner particles for gauge and Higgs bosons, which have to fulfill the corresponding naturalness sum rules.
On the other hand, other theories of naturalness, most notably supersymmetry, rely on partner particles with opposite statistics in comparison to their corresponding SM particles, yielding different kinematics.
Thus dedicated collider analyses are needed in order to explore the relevant tests of naturalness.
A full exploration of these possibilities will be deferred to future work.

\begin{acknowledgments}
J.~Hajer and T.~Liu would like to thank J.~Bernon, Y.-Y.~Li and J.~Shu for helpful discussions, while I.~Low acknowledges the hospitality at Hong Kong University of Science and Technology, where part of this work was performed.
J.~Hajer and T.~Liu are supported by the Collaborative Research Fund (CRF) under Grant \No{HUKST4/CRF/13G}.
T.~Liu is also supported by the General Research Fund (GRF) under Grant \No{16312716}.
Both the CRF and GRF grants are issued by the Research Grants Council of Hong Kong S.A.R..
H.~Zhang was supported by the U.S. DOE under Contract \No{DE-SC0011702}, as well as by IHEP under Contract \No{Y6515580U1} and IHEP Innovation Grant contract \No{Y4545171Y2}.
I.~Low was supported in part by the U.S. Department of Energy under contracts \No{DE-AC02-06CH11357} and \No{DE-SC 0010143}.
C.-R.~Chen was supported in part by the Ministry of Science and Technology of Taiwan under Grant \No{MOST-105-2112-M-003-010-MY3}.
\end{acknowledgments}

\appendix

\section{Higher Order Corrections}\label{sec:corrections}

In the main part of our discussion we have neglected higher order correction originating in the mass diagonalization after electroweak symmetry breaking.
In order to make good for this omission we add the following quartic terms to Lagrangian~\eqref{eq:gauge Lagrangian}
\begin{equation}
    \Delta \Lag_\GaugeL
  = \frac{\gauge_4}{4!f^3} \abs{H}^4 \gaugeR \GaugeL
  + \frac{\Gauge_4}{4!f^3} \abs{H}^4 \GaugeR \GaugeL
  + \text{h.c.}
\ .
\end{equation}
The rotation into mass eigenstates~\eqref{eq:mass rotation} leads to the following additional terms in Lagrangian~\eqref{eq:mass Lagrangian}
\begin{equation}
    \Delta \Lag_\MassL
  = \frac{\gamma_\massL}{4! m_\MassL^3} \abs{H}^4 \massR \MassL
  + \frac{\gamma_\MassL}{4! m_\MassL^3} \abs{H}^4 \MassR \MassL
  + \text{h.c.}
\ .
\end{equation}
The mass eigenstate parameter~\eqref{eq:parameter} at this order are in terms of gauge eigenstate parameter is given by
\begin{align}
    \gamma_\massL
 &= \left(\Gauge_0 \gauge_4 - \gauge_0 \Gauge_4 \right) \gauge^2
\ ,
  & \gamma_\MassL
 &= \left(\gauge_0 \gauge_4 + \Gauge_0 \Gauge_4 \right) \gauge^2
\ .
\end{align}
The rotation~\eqref{eq:electroweak rotation} of the right handed fermion fields up to cubic order in $v$ reads
\begin{subequations}
\begin{align}
    \breakR
 &= \massR
  - \MassR \frac{v^2}{2m_\MassL^2} (\alpha_\massL + 2 \lambda_\massL^* \lambda_\MassL)
  + \order{\frac{v^4}{m_\MassL^4}}
\ ,
 \\ \BreakR
 &= \MassR
  + \massR \frac{v^2}{2 m_\MassL^2} (\alpha_\massL + 2 \lambda_\MassL^* \lambda_\massL)
  + \order{\frac{v^4}{m_\MassL^4}}
\ .
\end{align}
\end{subequations}
At the same time the left handed fields are rotated by
\begin{subequations}
\begin{align}
    \breakL
=&\ \massL
  - \MassL \frac{v}{m_\MassL} \lambda_\MassL^*
  - \massL \frac{v^2}{2 m_\MassL^2} \abs{\lambda_\MassL}^2 \notag \\
 &- \MassL \frac{v^3}{6 m_\MassL^3}
    \left(
        \beta_\MassL^*
      + 3 \lambda_\massL^* \alpha_\massL
      - 3 \lambda_\MassL^* \left(\alpha_\MassL + \abs{\lambda_\MassL}^2 - 2 \abs{\lambda_\massL}^2 \right)
    \right)
  + \order{\frac{v^4}{m_\MassL^4}}
\ ,
 \\ \BreakL
=&\ \MassL
  + \massL \frac{v}{m_\MassL} \lambda_\MassL
  - \MassL \frac{v^2}{2m_\MassL^2} \abs{\lambda_\MassL}^2 \notag \\
 &+ \massL \frac{v^3}{6 m_\MassL^3}
    \left(
        \beta_\MassL
      + 3 \alpha_\massL \lambda_\massL
      - 3 \lambda_\MassL \left(\alpha_\MassL + \abs{\lambda_\MassL}^2 - 2 \abs{\lambda_\massL}^2\right)
    \right)
  + \order{\frac{v^4}{m_\MassL^4}}
\ .
\end{align}
\end{subequations}
The resulting quadratic correction to the top partner mass~\eqref{eq:partner mass} is
\begin{equation}
    m_\BreakL
  = m_\MassL
    \left(
        1
      + \frac{v^2}{2 m_\MassL^2} \left(\alpha_\MassL + \abs{\lambda_\MassL}^2\right)
      + \order{\frac{v^4}{m_\MassL^4}}
    \right)
\ .
\end{equation}
The Yukawa couplings~\eqref{eq:Yukawa couplings} are altered to
\begin{subequations}
\begin{align}
    \lambda_\breakL
 &= \lambda_\massL
  + \frac{v^2 }{2 m_\MassL^2}
     \left(
         \beta_\massL
       - 3 \left(
           \alpha_\massL \lambda_\MassL
         + \lambda_\massL \abs{\lambda_\MassL}^2
         \right)
     \right)
  + \order{\frac{v^4}{m_\MassL^4}}
\ ,
 \\ \lambda_\BreakL
 &= \lambda_\MassL
  + \frac{v^2}{2 m_\MassL^2}
    \left(
        \beta_\MassL
      + \alpha_\massL \lambda_\massL
      -  \lambda_\MassL
        \left(
            2 \left(\alpha_\MassL- \abs{\lambda_\massL}^2\right)
          + \abs{\lambda_\MassL}^2
        \right)
    \right)
  + \order{\frac{v^4}{m_\MassL^4}}
\ .
\end{align}
\end{subequations}
While the quadratic couplings~\eqref{eq:quadratic couplings} up to this order are
\begin{subequations}
\begin{align}
    \alpha_\breakL
 &= \alpha_\massL
  + \frac{v^2}{2 m_\MassL^2}
    \left(
        \gamma_\massL
      + 2
        \lambda_\MassL^*
        \left(
            \beta_\massL
          - \alpha_\MassL \lambda_\massL
        \right)
    \right)
  + \order{\frac{v^4}{m_\MassL^4}}
\ ,
 \\ \alpha_\BreakL
 &= \alpha_\MassL
  + \frac{v^2}{2 m_\MassL^2}
    \left(
        \gamma_\MassL
      + \alpha_\massL^2
      + \alpha_\MassL^2
      + 2 \left(\lambda_\MassL^*\beta_\MassL + \alpha_\massL \lambda_\massL^* \lambda_\MassL \right)
    \right)
  + \order{\frac{v^4}{m_\MassL^4}}
\ .
\end{align}
\end{subequations}
The correction to the VEV induced terms~\eqref{eq:electroweak Yukawa} up to quadratic order are
\begin{subequations}
\begin{align}
    \Delta \breakQuad_\breakL
 &= - \frac{v^2}{m_\MassL^2}
    \left(
        \frac{\gamma_\massL}{3}
      + \frac{\beta_\MassL^* \lambda_\massL}{3}
      + \lambda_\MassL^*
        \left(
            \beta_\massL
          - \alpha_\massL \lambda_\MassL
          - \lambda_\massL \left( \alpha_\MassL + \abs{\lambda_\MassL}^2 \right)
        \right)
    \right)
  + \order{\frac{v^4}{m_\MassL^4}}
\ ,
 \\ \Delta \breakQuad_\BreakL
 &= - \frac{v^2}{m_\MassL^2}
    \left(
        \frac{\gamma_\MassL}{3}
      + \frac{\beta_\MassL^* \lambda_\MassL}{3}
      + \lambda_\MassL^* (\beta_\MassL - 2 \alpha_\MassL \lambda_\MassL)
      - \abs{\lambda_\MassL}^4
    \right)
  + \order{\frac{v^4}{m_\MassL^4}}
\ .
\end{align}
\end{subequations}
We have refrained from writing down the quadratic corrections to the cubic couplings~\eqref{eq:cubic couplings} and~\eqref{eq:electroweak quadratic} as they involve terms of order $H^5$.

\bibliographystyle{JHEP}
\bibliography{references}

\end{document}